\documentclass[fleqn,usenatbib]{mnras}

\usepackage{newtxtext,newtxmath,hyperref}
\usepackage[T1]{fontenc}
\usepackage{graphicx} 
\usepackage{amsmath}	
\usepackage{amssymb}	
\usepackage{enumitem}
\usepackage[english]{babel}
\usepackage{gensymb}
\usepackage{changes}
\usepackage{color, colortbl}

\definechangesauthor[name={NT}, color=orange]{NT}
\setremarkmarkup{(#2)}

\newcommand{\lpos}{0.004} 
\newcommand{\mpos}{0.02} 
\newcommand{\hpos}{0.1} 
\newcommand{\lbeam}{0.02} 
\newcommand{\mbeam}{0.1} 
\newcommand{\hbeam}{0.5} 
\newcommand{\lpoint}{0.04} 
\newcommand{\mpoint}{0.2} 
\newcommand{\hpoint}{1.0} 

\newcommand{\Real}[1]{\mathbb{R}\text{e}\hspace{-1pt}\left[#1\right]}
\newcommand{\Imag}[1]{\mathbb{I}\text{m}\hspace{-1pt}\left[#1\right]}

\makeatletter
\def\@fnsymbol#1{\ensuremath{\ifcase#1\or {}^\star \or {}^{\star\star}\or \dagger\or \ddagger\or
    \mathsection\or \mathparagraph\or \|\or **\or \dagger\dagger
    \or \ddagger\ddagger \else\@ctrerr\fi}}
\makeatother

\title[Mitigating Redundant Calibration Errors]{Mitigating the Effects of Antenna-to-Antenna Variation on Redundant-Baseline Calibration for 21\,cm Cosmology}

\author[N. Orosz et al.]{Naomi Orosz$^{1}$\thanks{Email: naomiorosz@berkeley.edu},
Joshua S. Dillon$^{1\dagger}$\thanks{Email: jsdillon@berkeley.edu},
Aaron Ewall-Wice$^{2}$,
Aaron R. Parsons,$^{1}$ \newauthor
Nithyanandan Thyagarajan$^{3,4,\ddagger}$
\\
$^{1}$Department of Astronomy, University of California, Berkeley, Berkeley, CA, USA\\
$^{2}$ Jet Propulsion Laboratory, California Institute of Technology, Pasadena, CA, USA\\
$^{3}$ National Radio Astronomy Observatory, Socorro, NM, USA\\
$^{4}$ Arizona State University, School of Earth and Space Exploration, Tempe, AZ, USA \\
$^{\dagger}$NSF Astronomy and Astrophysics Postdoctoral Fellow\\
$^{\ddagger}$Jansky Postdoctoral Fellow\\
}

\date{\today}
\pubyear{2017}

\begin{document}
\label{firstpage}
\pagerange{\pageref{firstpage}--\pageref{lastpage}}
\maketitle

\begin{abstract}
The separation of cosmological signal from astrophysical foregrounds is a fundamental challenge for any effort to probe the evolution of neutral hydrogen during the Cosmic Dawn and epoch of reionization (EoR) using the 21\,cm hyperfine transition. Foreground separation is made possible by their intrinsic spectral smoothness, making them distinguishable from spectrally complex cosmological signal even though they are $\sim$5 orders of magnitude brighter. Precisely calibrated radio interferometers are essential to maintaining the smoothness and thus separability of the foregrounds. One powerful calibration strategy is to use redundant measurements between pairs of antennas with the same physical separation in order to solve for each antenna's spectral response without reference to a sky model. This strategy is being employed by the Hydrogen Epoch of Reionization Array (HERA), a large radio telescope in South Africa that is now observing while being built out to 350 14-m dishes. However, the deviations from perfect redundancy inherent in any real radio telescope complicate the calibration problem. Using simulations of HERA, we show how calibration with antenna-to-antenna variations in dish construction and placement generally lead to spectral structure in otherwise smooth foregrounds that significantly reduces the number of cosmological modes available to a 21\,cm measurement. However, we also show that this effect can be largely eliminated by a modified redundant-baseline calibration strategy that relies predominantly on short baselines. 
\end{abstract}



\section{Introduction} 

The 21\,cm spin transition line of neutral hydrogen has the potential to become a powerful probe of the EoR, complementing limits on the end of reionization from quasars and its duration from the cosmic microwave background. By measuring temperature, ionization, density fluctuations across cosmic time, 21\,cm cosmology may in time become a highly sensitive tool for understanding our universe from the Dark Ages and the Cosmic Dawn to cosmology and fundamental physics \citep{FurlanettoReview, miguelreview, PritchardLoebReview, SaleemEoRChapter, aviBook, MesingerBook}.

Advancements in instrumental precision, sensitivity, and redshift coverage have given rise to the numerous experiments working to detect the 21\,cm signal over a wide range of redshifts, both through spatial fluctuations in brightness temperature and the angularly averaged ``global'' signal. While many experiments have focused on the impact of the EoR on the 21\,cm brightness temperature, an earlier period of the Cosmic Dawn during which the first stars and X-ray sources turned on has recently garnered significant interest due to the (as yet unconfirmed) detection of a strong global 21\,cm absorption feature by the Experiment to Detect the Global Epoch of Reionization Signature (EDGES; \citealt{EDGES}). Because its magnitude exceeds previous predictions by at least a factor of two, it has spurred a number of interesting interpretations, including the existence an unknown population of high redshift radio sources \citep{FengHolderARCADE,EwallWiceRadioBackground} or the presence of baryon-dark matter interactions (e.g.\ \citealt{EDGES-darkmatter}).

That said, all efforts to detect cosmological hydrogen via the 21\,cm line are complicated by overwhelmingly bright astrophysical foregrounds. These foregrounds, which are determined by synchrotron radiation from our Galaxy and nearby galaxies but also include thermal Bremsstrahlung from H II regions, are $\sim$10$^5$ times brighter than the cosmological signal. However, because of their smooth spectral structure, they occupy the lowest Fourier modes along the line of sight in the power spectrum of fluctuations. When observed by interferometers, which are inherently chromatic, intrinsically smooth foregrounds are spread out in 2D Fourier space\footnote{$k_\| $ describes cosmological modes along the line of sight and $k_\perp$ combines modes in the $x-$ and $y$ directions (i.e.\ perpendicular to the line of sight). Larger $k_\|$ modes corresponding to finer Fourier modes in the frequency response; larger $k_\perp$ modes correspond to longer baselines.} into a ``wedge''-shaped region \citep{Dattapowerspec,AaronDelay,VedanthamWedge,MoralesPSShapes,Hazelton2013,PoberWedge,ThyagarajanWedge,NithyaPitchfork,EoRWindow1,EoRWindow2}. The rest of Fourier space, which is nominally foreground-free, is called the ``EoR window,'' though the concept is equally valid for higher and lower redshift 21\,cm observations. The boundary between the wedge and the window depends on $k_\perp$ (and thus baseline length) and is given by the light travel time delay of a source at the horizon between two elements \citep{AaronDelay}. Filtering out all foregrounds and signal within the wedge makes a robust 21\,cm detection still possible in the window, albeit at the cost of sensitivity \citep{PoberNextGen}; working within the wedge requires extremely precise instrument modeling and foreground subtraction. With such bright astrophysical foregrounds, precise calibration becomes essential to keeping the EoR window clean enough to actually detect the cosmological signal. If by miscalibrating we introduce chromatic errors, we can no longer rely on the assumption that foregrounds stay in the wedge even if our instrument has a smooth spectral response \citep{MoralesPSShapes, X13, YatawattaConsensus, BarryCal, ModelingNoise}.

The problem of calibration\footnote{By calibration, we mean bandpass calibration of a single complex number per frequency and per antenna, not ``direction-dependent calibration'' of antenna primary beams. In this paper, we also ignore the complexities of polarization \citep{DillonPolarizedRedundant} and polarization leakage via $D$-terms.\citep{RadioPolarimetry1, RadioPolarimetry2}.} boils down to one equation:
\begin{equation} \label{eq:gains}
    V_{ij}^\text{obs}(\nu) = g_i(\nu)g_j^*(\nu)V_{ij}^\text{true}(\nu) + n_{ij}(\nu),
\end{equation}
where $V_{ij}^\text{true}$ is the inherent visibility of the baseline between antennas $i$ and $j$, $g_i$ and $g_j$ are the two complex gains for each antenna, $n_{ij}$ is the noise, and $V_{ij}^{\text{obs}}$ is the visibility actually measured by the instrument \citep{ThompsonMoranSwenson}. How one approaches calibration depends on which specific terms are treated as known \textit{a priori} and thus which terms need to be solved for. Sky-based self-calibration algorithms start with a model of the radio sky and antenna beams to produce a best guess for the true visibility \citep{selfcal,selfcalreview}. This allows one to solve for the antenna gains using the observed data. The Giant Metrewave Radio Telescope \citep{GMRT}, the Low-Frequency Array \citep{LOFAR, PatilLOFARLimit}, and the Murchison Widefield Array (MWA; \citealt{MWA, EmpiricalCovariance,AaronFirstEoXLimits,CHIPS,BeardsleyFirstSeason}) employ variations of this method.

Though sky-based calibration is popular and well-suited to arrays optimized for interferometric imaging, it has significant limitations in the context of 21\,cm observations. It depends on both the accuracy and depth of existing catalogue and survey data. The enormous dynamic range between foregrounds and cosmological signal means that even relatively faint sources now collectively pose a problem \citep{PatilLOFARCal}. Simulations by \cite{BarryCal} indicate that traditional sky-based calibration requires unprecedented catalogue depth and accuracy to prevent these faint, unmodeled sources from introducing chromatic visibility errors that mix foregrounds into otherwise clean Fourier modes. This occurs because these sources, although faint, are numerous and contribute significantly to the total flux of any given field, introducing modeling errors with the same chromaticity as the visibilities themselves, which then affect all visibilities via gain calibration errors. \cite{BarryCal} examine different solutions, such as fitting the bandpass with low-order polynomial to ensure spectral smoothness and averaging calibration solutions over antennas and/or time.  \cite{ModelingNoise} explained this effect analytically as a form of chromatic noise in the power spectrum which does not average down in time, even when the source model contains accurate fluxes well below the confusion limit. However, they also showed that re-weighting data before calibration to prioritize short baselines (i.e.\ baselines with less spectral structure in their visibilities) helped recover most of the EoR window.

Short of developing external calibration systems or making orders-of-magnitude improvements in beam and source modelling, one alternative approach is to calibrate using the internal consistency of redundant visibility measurements \citep{redundant, Wieringa}. Redundant-baseline calibration starts with many measurements of $V_{ij}^{\text{obs}}$ from a group of baselines that measure the true visibilities because they share a common baseline vector and putatively identical primary beams. This redundancy allows one to solve, up to a handful of degeneracies, for gains and true visibilities without assuming the sky model  \citep{MITEoR, ZhengBruteForce, DillonPolarizedRedundant}. Since baselines belonging to the same redundant group should all see the same inherent visibility, $V_{ij}^{\text{true}}$, multiplied by the antenna gains involved in the specific baseline, a least-squares estimator can solve for the true visibility and antenna gains that reproduce the measured visibilities. This approach is being used by HERA \citep{RedArrayConfig,HERAOverview}, as well predecessor telescopes like the MITEoR experiment \citep{MITEoR}, the Donald C.\ Backer Precision Array for Probing the Epoch of Reionization \citep{PAPER-64}, and Phase II of the MWA \citep{LiMWAIICalibration}.

As with any calibration method, redundant-baseline calibration comes with limitations as well as system design requirements. Its explicit dependence on the redundancy of measurements means that the antenna responses of the baselines within a redundant group must be identical as well. We know from \cite{BarryCal} that sky modeling errors produce chromatic calibration errors that contaminate the EoR window. One naturally wonders: do analogous errors arise due to the assumption of redundancy in a not-quite-redundant array? And, if so, can they be mitigated using similar down-weighting strategies to those proposed by \cite{ModelingNoise}?

In this work, we simulate the visibilities observed by a not-quite-redundant version of the HERA core by introducing antenna position errors and beam-to-beam variations at various realistic levels. In Section~\ref{sec:simulations} we detail our simulations and show the effects of non-redundancy on the 21\,cm power spectrum after redundant-baseline calibration. Then, in Section~\ref{sec:mitigating}, we adapt the strategy of \cite{ModelingNoise} for mitigating the effects of redundancy errors on the power spectra through a baseline-length-dependent reweighting of the data.


\section{Simulations of Redundancy Errors and the Effect on the Power Spectrum} \label{sec:simulations}

In order to understand the effect of non-redundancies, we need to simulate an array where the true gains and visibilities are known. 
We start by explaining in Section~\ref{sec:vissim} how we simulate visibilities and parameterize various forms of non-redundancy.
Next in Section~\ref{sec:redcal} we review how redundant-baseline calibration works and examine in Section~\ref{sec:non-redcal} its qualitative effects when non-redundancies are introduced. Then in Section~\ref{sec:nonred} we demonstrate how these non-redundancies show up in power spectra.

\subsection{Visibility simulation with non-redundancy}
\label{sec:vissim}

In an ideal interferometer, the correlation between the electric fields measured by antennas $i$ and $j$ is a \textit{visibility}, a particular angular Fourier mode of $I(\mathbf{\hat{r}},\nu)$, the specific intensity of the sky, as a function of direction and frequency. Ignoring noise, it is given by 
\begin{equation} \label{eq:truevis}
    V^\text{true}_{ij}(\nu) = \int B_{ij}(\mathbf{\hat{r}}, \nu) I(\mathbf{\hat{r}},\nu) \exp \left[-2\pi i \frac{\nu}{c}\mathbf{b}_{ij} \cdot \mathbf{\hat{r}} \right] d\Omega,
\end{equation}
where $B_{ij}$ is the primary beam and $\mathbf{b}_{ij}$ is the \textit{baseline}, the vector separation between antennas $i$ and $j$ \citep{ThompsonMoranSwenson}. For simplicity, we simulate visibilities from a set of point sources rather than a whole-sky integral, approximating Equation~\ref{eq:truevis} as
\begin{equation} \label{eq:vissum}
    V^\text{true}_{ij}(\nu) \approx \sum_{n = 1}^{N_\text{sources}} B_{ij}(\mathbf{\hat{r}}_n, \nu) S_{n}(\nu) \exp \left[-2\pi i \frac{\nu}{c}\mathbf{b}_{ij} \cdot \mathbf{\hat{r}}_n \right]
\end{equation}
where $n$ indexes over the sources, their fluxes $S_n$, and their positions $\mathbf{\hat{r}}_n$.

In this paper, we adopt a fairly simple sky model in order to isolate the effects of non-redundancy on redundant calibration. We take all sources from the GaLactic and Extragalactic All-sky MWA (GLEAM) survey \citep{GLEAM} above 15\,Jy  at 151\,MHz, for a total of 126 total sources with well-measured fluxes and spectral indices. Because we are primarily interested in calibration using redundant baselines, the fidelity of our model sky to the true radio sky is not as important as it was in \citet{BarryCal} and \citet{ModelingNoise}. Rather, we need enough sources spread across the sky to produce realistic spectral structure in our simulated visibilities using Equation~\ref{eq:vissum}.\footnote{The fact the wedge is fairly uniformly filled at different delays in the Airy beam panel of Figure \ref{fig:no_errors} validates this assumption.} To produce a frequency range and spectral resolution suitable for power spectrum estimation \citep{yi, PAPER32Limits}, we simulate our sources in 100 frequency channels from 140\,MHz to 160\,MHz.

\begin{figure}
\includegraphics[width = \linewidth]{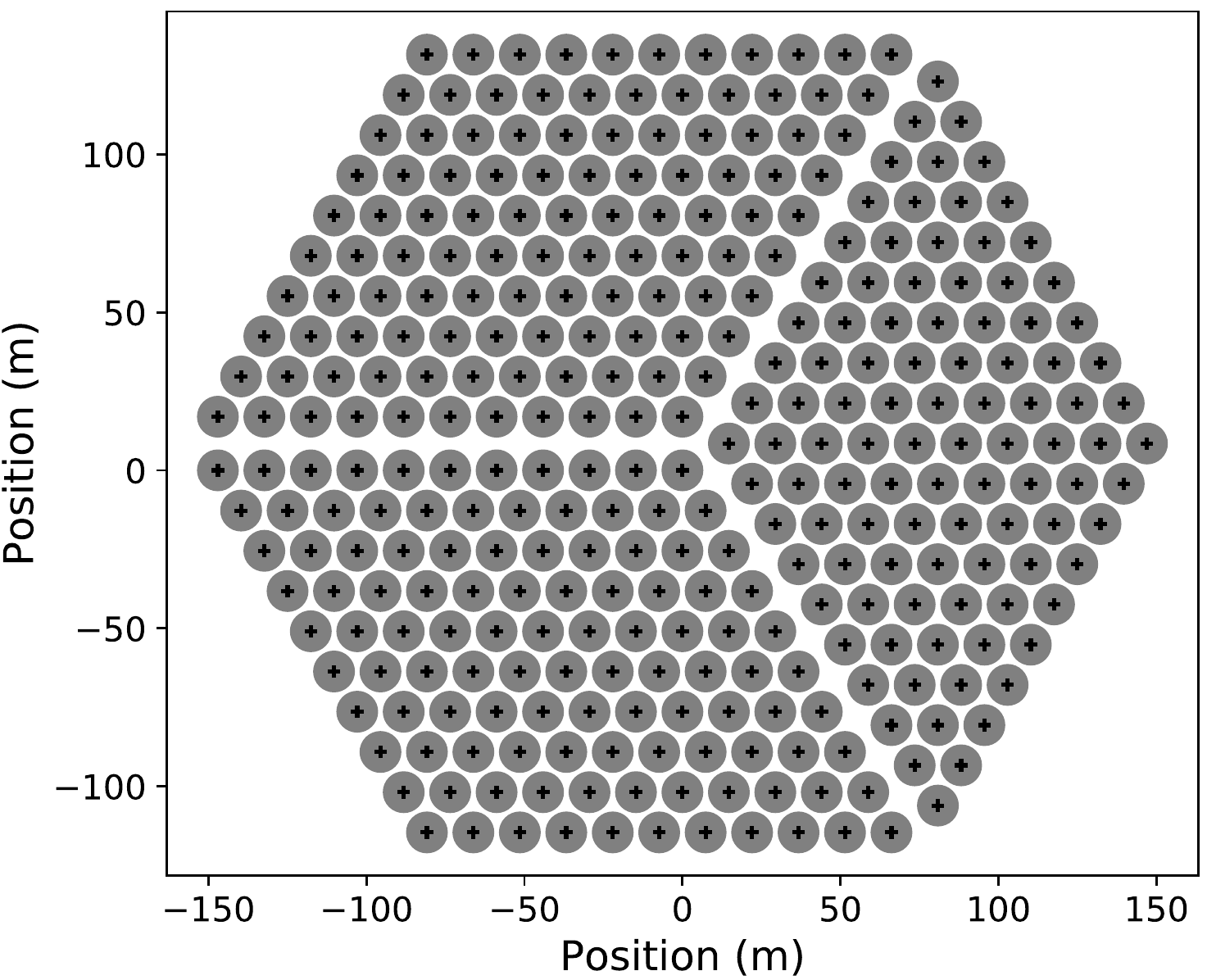}
\caption{HERA's 320-antenna core features a dense, hexagonal packing with 14.6\,m baselines. The dense packing maximizes sensitivity on the short baselines that are the least contaminated by the wedge. The split into three slightly offset sectors allows it to sample nearly triple the number of unique instantaneous baselines while still being redundantly calibratable \citep{RedArrayConfig}. The 30 planned outrigger dishes, which increase HERA's angular resolution, are excluded from our simulations. While they are potentially redundantly calibratable, their exclusion speeds up simulations and calibrations considerably and does not significantly affect our conclusions.}
\label{fig:layout}
\end{figure}

For our array configuration, we adopt the core of the HERA array layout as our prototypical worked example. HERA's core includes 320 dishes arranged in a hexagon split into three sectors (Figure~\ref{fig:layout}). This split nearly triples the number of unique baselines sampled while still being redundantly-calibratable \citep{RedArrayConfig}. The centers of the 14-m meter dishes are generally separated by only 14.6\,m, resulting in a dense packing that maximizes sensitivity on short baselines, where the cosmological signal is the strongest outside the wedge (i.e. where foregrounds are most compact in cosmological Fourier space). HERA is located at -30.722$^{\circ}$ latitude and, as a drift-scan instrument, observes a stripe of constant declination. We simulate zenith-pointed observations centered at $\text{RA} = 10$\,h and $\text{DEC} = -30.722^{\circ}$, though the calibration methods developed in this paper do not depend on a specific field. Our simulations are free of thermal noise in order to attain the high dynamic range we need study foreground bias in the EoR window of comparable magnitude to the cosmological signal.

As a simple beam model that still exhibits complex, all-sky spatial structure, we adopt an Airy function as our primary beam. Our zenith-pointed Airy beam takes the form
\begin{equation}\label{eq:airy}
    B_{ij}(\mathbf{\hat{r}}_n, \nu) = \left[\frac{2J_1(ka\sin\theta)}{ka\sin\theta}\right]^2,
\end{equation}
where $\theta$ is the co-altitude, $J_1$ is the Bessel function of the first kind of order one, $k=\frac{2\pi\nu}{c}$ is the wave number, and the aperture radius is $a = 6\,$m to account for under-illumination of the HERA dish \citep{HERAOverview}. This produces a beam with a full-width at half-maximum (FWHM) is $1.03c / 2a\nu $, which is 9.8$^\circ$ at 150\,MHz. Though there are important differences in their detailed shapes, the Airy beam features spectral and spatial structure that is broadly similar to that of real HERA beams \citep{NebenHERADish1}.
In Section~\ref{sec:nonred} we also examine a Gaussian beam which proved too simplistic to capture the effects of non-redundancy on the power spectrum. 

In this paper, we investigate three sources of non-redundancy due to deviations from ideal antenna elements that are sited perfectly and constructed identically. These are: 1) antenna position errors, 2) beam size/shape errors, and 3) beam pointing errors. While these do not entirely capture the diversity of antenna-to-antenna variations in a real array, they constitute a reasonable basis for the kind of non-idealities in element and feed placement and orientation that are, at some level, inevitable. We investigate their impact on redundant calibration by parameterizing each of them at three different levels spanning a realistic range. 

Ideal antennas would be situated at their precise assigned grid points with a zenith-pointed, radially-symmetric beam with a FWHM of 9.8$^\circ$ at 150\,MHz. We simulate non-redundancy by introducing mean-0, normally-distributed variations in all three sources of error in both the $x$- and $y$-directions. We generalize the Airy function (Equation~\ref{eq:airy}) to represent a beam that can vary in both its shape and pointing:
\begin{equation}\label{eq:airy_general}
    B_{i}(\mathbf{\hat{r}}, \nu) = \left[ \frac{2J_1\left(k\sqrt{a_x^2(x-x_{pc})^2 + a_y^2(y-y_{pc})^2}\right)}{k\sqrt{a_x^2(x-x_{pc})^2 + a_y^2(y-y_{pc})^2}} \right]^2,
\end{equation} 
and we model the beam for a given baseline as $B_{ij} = \sqrt{B_iB_j}$ to account for the individual contributions of the two antennas that form the baseline. 
Pointing errors are represented by randomly picking a pointing center $\mathbf{\hat{r}}_{pc}$ near but generally not at zenith (i.e.\ $x_{pc}$ and $y_{pc}$ are normally-distributed with mean 0). Likewise, beam size and shape errors are expressed by independently varying $a_x$ and $a_y$ around the mean of value of 6\,m to alter the semi-major and semi-minor axes of the aperture. In each simulation, we apply statistically independent errors to each antenna and then simulate all visibilities for all antenna pairs.

\begin{figure*}
\includegraphics[width = \textwidth]{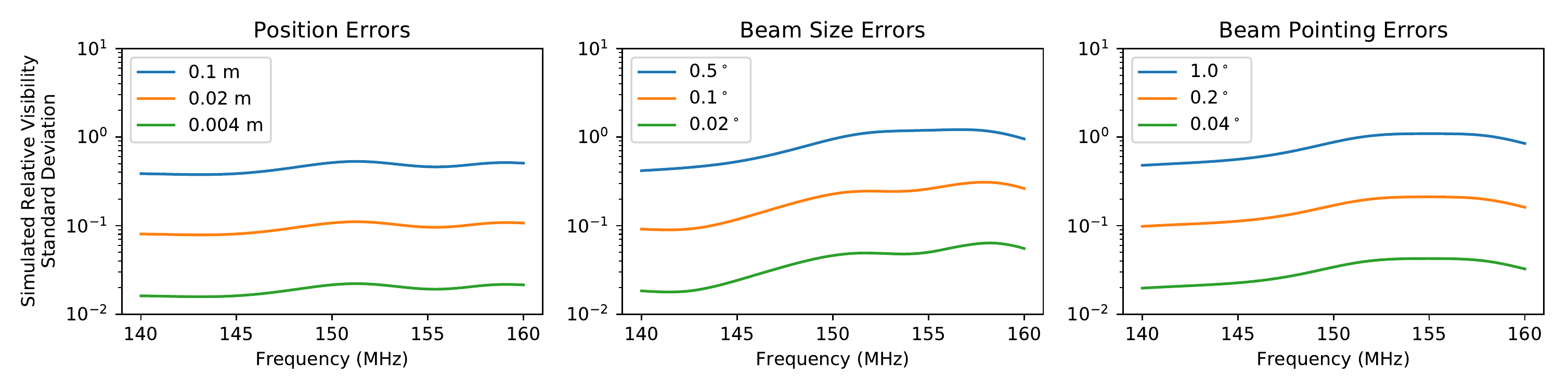}
\caption{Here we show the ratio of the standard deviations of all the simulated visibilities to the error-free visibility for a single unique 14.6\,m baseline for all three redundancy error types, each with low, fiducial, and high error levels. Error levels were selected both to represent plausible ranges of redundancy errors in HERA and to produce standard deviations with comparable magnitudes.}
\label{fig:STDEV}
\end{figure*}

We choose low, fiducial, and high error levels for each non-redundancy type, which we list in Table \ref{errors_table}. 
\begin{table}
\centering
    \begin{tabular}{|c||c|c|c|}
         \hline
            & Position & Beam Size & Beam Pointing \\
        \hline
         Low & \lpos \space m & \lbeam $^{\circ}$ & \lpoint $^{\circ}$ \\
        \hline
        Fiducial & \mpos \space  m & \mbeam $^{\circ}$ & \mpoint $^{\circ}$  \\
        \hline
        High & \hpos \space  m & \hbeam $^{\circ}$ &\hpoint $^{\circ}$ \\
        \hline
    \end{tabular}
    \caption{Low, fiducial, and high error values for position, beam size, and beam pointing errors span the range of realistic levels of non-redundancy in HERA. Position errors come from antenna and dish placement, beam size errors encompass variations feed height and dish diameter, and beam pointing errors occur due to variations in feed placement and orientation.}
    \label{errors_table}
\end{table}
These were chosen to represent realistic\footnote{HERA's large collecting area is enabled by employing relatively inexpensive materials \citep{HERAOverview}, limiting the repeatability of antenna construction. HERA's antenna placement and feed positioning are accurate to within 2\,cm \citetext{D.~R.~DeBoer, Private Communication, 2018}. This sets our fiducal error level. At HERA's 4.5\,cm focal height, this would correspond to $\arcsin{\left(\text{2\,cm}/\text{4.5\,m}\right)} = 0.25^\circ$ pointing errors. Likewise, HERA's FWHM decreases roughly $2^\circ$ for every meter of feed height \citep{NebenHERADish1}, which would imply $0.04^\circ$ beam size errors. We take a more conservative value of $0.1^\circ$ to reflect differences in dish construction.} antenna-to-antenna variations \citep{CarilliRedundantClosure}, with the condition that the resulting errors in visibilities are comparable for sources of non-redundancy at the same error level. To verify this, we show in Figure~\ref{fig:STDEV} the standard deviation of all the nominally redundant measurements of a single unique 14.6\,m baseline.
The low, fiducial, and high error levels produce visibility variations on the $\sim$1\%, $\sim$10\%, and $\sim$100\% levels, respectively, spanning the range of plausible levels of non-redundancy in HERA.

\subsection{A review of redundant baseline calibration}
\label{sec:redcal}

In order to solve for gains and visibilities, Equation~\ref{eq:gains} can be written in a system of equations in which each instance of the equation describes the visibility measured by a baseline. The unknowns in the system of equations are the complex gains (one per antenna) and the visibility solutions. However, since HERA is highly redundant, there are far fewer unique visibility solutions than measured visibilities (and thus equations). HERA's core, as shown Figure~\ref{fig:layout}, has 51,040 total baselines but only 1,501 unique baselines. 
Assuming antennas have the same beam, two baselines with the same displacement measure the same modes on the sky, up to the product of their antennas' complex gains. The result of this redundancy between baselines is an overdetermined non-linear system of equations for which we would like to minimize $\chi^2$, defined as
\begin{equation}\label{eq:chi_squared}
    \chi^2 = \sum\limits_{\text{all pairs } i,j}{}\frac{\left|V_{ij}^\text{obs}-g_i(\nu)g_j^*(\nu)V_{i-j}^\text{sol}(\nu)\right|^2}{\sigma_{ij}^2(\nu)},
\end{equation}
where $\sigma_{ij}^2(\nu)$ is the variance of $n_{ij}(\nu)$, the random Gaussian noise on visibilities,\footnote{Though our simulations do not include thermal noise, we minimize $\chi^2$ as if $\sigma_{ij}$ were constant for all pairs $i$ and $j$.} and $V_{i-j}^\text{sol}$ is the visibility solution for all baselines with the same separation as $V_{ij}$. Due to the form of $\chi^2$, there are four degeneracies that remain in the visibility and gains solutions (in single-polarization calibration; \citealt{redundant, DillonPolarizedRedundant}). The simplest of these, the overall amplitude, arises from the fact that if one gain is multiplied by a constant and the visibility solution is divided by the square of that constant, $g_i(\nu)g_j^*(\nu)V_{i-j}^\text{sol}(\nu)$ remains unchanged. The other degeneracies can be interpreted as potential changes to the phases of gains and visibilities that also cancel out for all baselines. 

One approach to minimizing $\chi^2$ is \texttt{lincal}, originally developed in \cite{redundant}. \texttt{lincal} takes an initial guess for gains and visibilities and iteratively minimizes $\chi^2$ using the Gauss-Newton algorithm. Equation~\ref{eq:gains} can be expanded around starting gains $g_i^0$ and visibilities $V_{i-j}^0$ and written as
\begin{equation}\label{eq:expanded_vis}
   V_{ij}^\text{obs} = (g_i^0 + \Delta g_i)(g_j^0 + \Delta g_j)^*(V_{i-j}^0 + \Delta V_{i-j}) , 
\end{equation}
where the $\Delta$ terms are solved for at each iteration, allowing the input guesses to be updated for the next iteration. Assuming the initial guess is close, the $\Delta^2$ terms are negligible and the equation becomes linear:
\begin{equation}\label{eq:expanded_vis2}
    V_{ij}^\text{obs}-g_i^0g_j^{0*}V_{i-j}^0 = \Delta g_ig_j^{0*}V_{i-j}^0 + g_i^0\Delta g_j^*V_{i-j}^0 + g_i^0g_j^{0*} \Delta V_{i-j}. 
\end{equation}

Because Equation~\ref{eq:expanded_vis2} contains complex conjugation, it must be broken into real and imaginary parts for the system of equations to be written as a matrix and solved with standard linear algebra techniques. Thus
\begin{align}\label{eq:Re}
  \mathbb{R}\text{e} \left[V_{ij}^\text{obs}\right. & \left. - \mbox{ } g_i^0g_j^{0*}V_{i-j}^0 \right] = \\ \nonumber
  &\Real{\Delta g_i} \Real{g_j^{0*}V_{i-j}^0} - \Imag{\Delta g_i}\Imag{g_j^{0*}V_{i-j}^0} + \\ \nonumber
  &\Real{\Delta g_j}\Real{g_i^0V_{i-j}^0} + \Imag{\Delta g_j}\Imag{ g_i^0V_{i-j}^0} + \\ \nonumber
  &\Real{\Delta V_{i-j}}\Real{g_i^0g_j^{0*}} - \Imag{\Delta V_{i-j}}\Imag{g_i^0g_j^{0*}} 
\end{align}
and analogously for the imaginary part.

This system can be expressed as the matrix equation
\begin{equation}
    \mathbf{d} = \mathbf{Ax},
\end{equation}
where $\mathbf{d}$ contains the left-hand side of Equation~\ref{eq:Re} and its imaginary counterpart and has a length equal to twice the number of observed visibilities, $\mathbf{x}$ contains all the real and imaginary $\Delta$ terms to be solved for, and $\mathbf{A}$ contains their coefficients. Because $\mathbf{x}$ contains the real and imaginary parts of the variables (gains and visibilities), it has a length of twice the number of unique visibilities plus twice the number of antennas. Likewise, the number of columns in $\mathbf{A}$ is equal to twice the number of true visibilities plus twice the number of gains and the number of rows in $\mathbf{A}$ is twice the number the measured visibilities. 

We can form a linear estimator for the $\Delta$ terms in $\mathbf{x}$ of the form
\begin{equation} \label{eq:x_with_weights}
    \mathbf{\hat{x}} = \left(\mathbf{A}^\intercal \mathbf{W}\mathbf{A} \right)^{-1} \mathbf{A}^\intercal \mathbf{W} \mathbf{d}
\end{equation}
where $\mathbf{W}$ is a matrix of data weights. If the only source of error were noise, the optimal $\mathbf{W}$ (in a least-squares sense) would be $\mathbf{N}^{-1}$ where $\mathbf{N}$ is the noise covariance of the data. As we will discuss in Section~\ref{sec:mitigating}, non-redundancy motivates us to consider alternative weighting schemes following \cite{ModelingNoise}.

\subsection{Redundant baseline calibration with non-redundancy}
\label{sec:non-redcal}
Redundant calibration depends on the assumption that, up to a complex, per-antenna gain, baselines with the same displacement measure the same modes on the sky and therefore there are far fewer unique true visibilities than observed visibilities. Due to position errors and variations in the beam size and pointing, true visibilities are not redundant even after calibration. Deviations from redundancy do not necessarily preclude redundant baseline calibration, but rather act like a source of variation that leads to gain errors.  

When performing redundant baseline calibration with non-redundancy, the gains and visibilities that minimize $\chi^2$ are not generally the ``correct'' gains and visibilities, i.e.\ the result of a simulation without redundancy errors. However, as long as baselines are still approximately redundant, \texttt{lincal} can be used to minimize $\chi^2$. In our simulations, we start with the ``correct'' gains and visibilities as the initial guess then use the iterative least-squares estimator described above to move toward the gain and visibility solutions that minimize $\chi^2$. In this way, we avoid confusing the effects of a failure to find the minimum $\chi^2$ with spectral structure on the gains in the case where $\chi^2$ is successfully minimized. For our \textit{absolute calibration}, a second step is necessary to fix degeneracies in amplitude and phase inherent to redundant-baseline calibration. We simply adopt the ``correct'' values within the degenerate subspace. This ensures that the degeneracies, which emerge from the structure of $\chi^2$, do not affect the final answer, enabling a more direct comparison between solutions. The problem of absolute calibration (which generally requires a sky model) in the presence of non-redundancy is left for future work (e.g. \citealt{RubyAbscal}).

Because gains and visibilities are solved for in the same system of equations, chromatic errors in the true visibilities can affect the estimates of the gains in a frequency-dependent way. Visibility errors on longer baselines can lead to visibility errors on shorter baselines via gain errors on shared antennas. Longer baselines exhibit more intrinsic spectral structure (this is, after all, the origin of the wedge) and any leakage of structure could contaminate previously clean modes on short baselines. We might therefore worry that chromatic gain errors due to miscalibration could lead to foreground bias that contaminates the EoR window, as was observed in \cite{BarryCal} and \cite{ModelingNoise}.

\begin{figure}
\includegraphics[width = \linewidth]{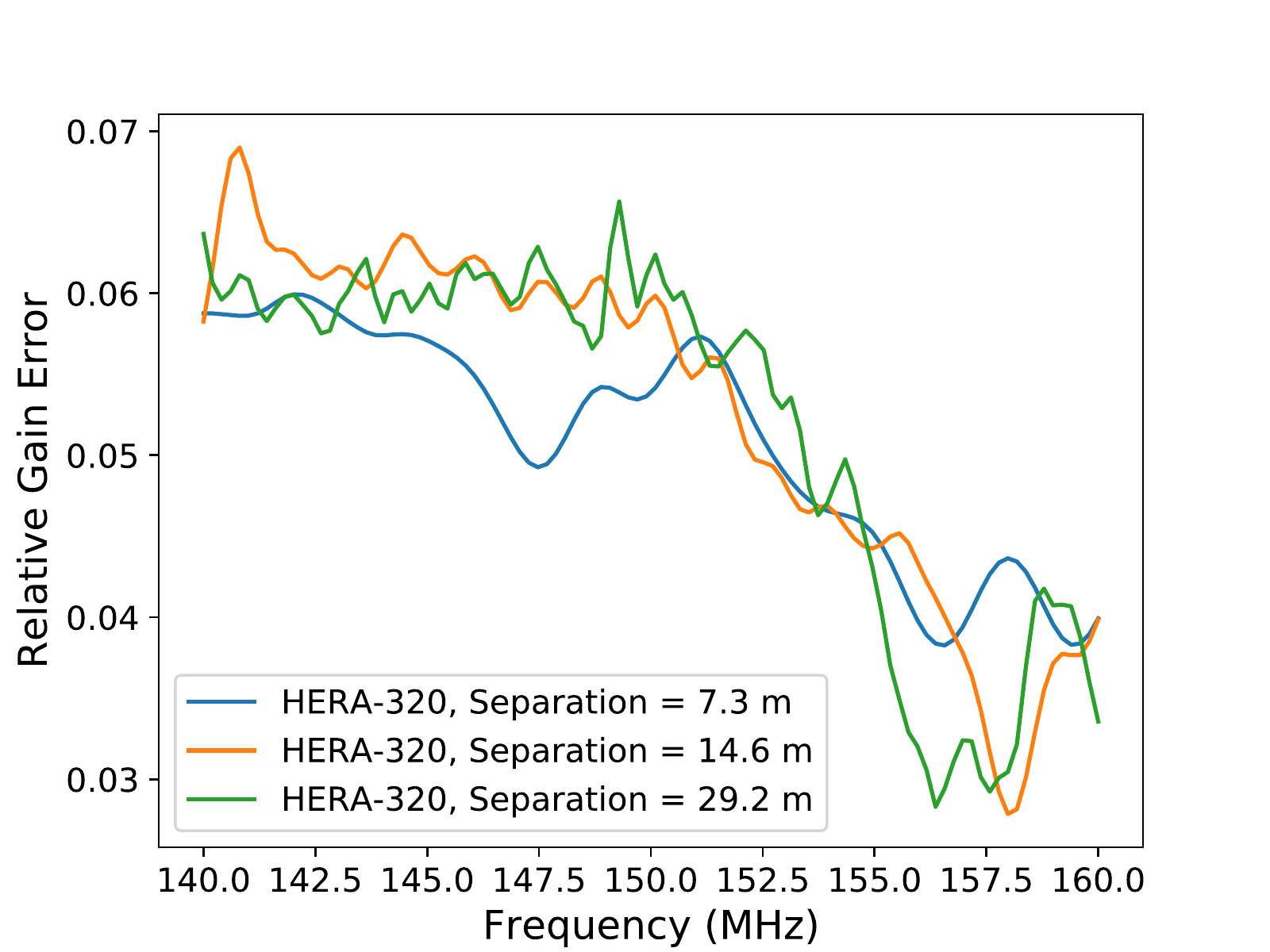}
\caption{The relative difference between calibrated gain solutions and true gains for a single antenna in the Northwest corner of the array is shown in arrays with one-half, one, and two times the normal separation between antennas, all with fiducial errors and otherwise identical. Arrays with longer baselines show more spectral structure in the gain solutions. Since all antennas are part of long and short baselines and these baselines are interlinked in calibration, spectral structure from the long baselines affects all antennas. This qualitative trend is true for antennas across the array.}
\label{fig:gains}
\end{figure}

We begin our investigation of this effect in the context of redundant-baseline calibration with non-redundancies by qualitatively examining the spectral structure in gain errors. We simulate true visibilities with our sky model and antennas, each with their random position and beam errors. To these, we apply simulated gains to represent the analog signal chain, each with a amplitude and delay of the form
\begin{equation}
    g_i(\nu) = e^{\gamma_i + 2\pi i \tau_i \nu},
\end{equation}
where $\gamma_i$ is normally distributed around 0 with standard deviation 0.2 and the delay $\tau_i$ is normally distributed around 0 with standard deviation 0.5\,ns. This is meant to represent a reasonable error after delay calibration while not introducing any instrumental spectral structure that we would later have to tease apart from spectral structure due to calibration errors.  Figure~\ref{fig:gains} shows the difference between calibrated gain solutions and the true simulated gains for a single representative antenna as a function of frequency.

The spectral smoothness of the intrinsic gains is an important caveat for this work. In our numerical experiments with artificial ``white noise'' in the simulated gains, we found that intrinsic small-scale spectral structure at greater than the 0.1\% level can have a major impact on the EoR window. An exploration of non-redundancy in the context of gains with realistic spectral structure is left for future work.

To investigate how gain errors change with overall baseline length, we simulate arrays with one-half and twice the antenna separation of the standard HERA core (Figure~\ref{fig:gains}). Although an array with half the standard antenna separation (7.3\,m) is physically impossible because dishes would overlap, we include it to illustrate the effect of a relative lack of long baselines. The gain errors in the largest array (separation 29.2\,m) shows spectral structure on finer scales than the standard HERA core, and the half-separation array displays the least spectral structure. Due to the densely-packed arrangement of HERA, all antennas involved in long baselines also participate in short baselines, allowing chromatic errors from long baselines to spread to short ones during calibration. Larger separations means longer baselines, producing more spectral structure that leaks power to fine scales. This is particularly worrisome because chromatic gain errors lead to spectral structure in calibrated visibilities for short baselines, which can leak power outside the wedge. We now turn to a more quantitative examination of that effect on the 21\,cm power spectrum.

\subsection{Effects of non-redundancy on power spectra}
\label{sec:nonred}
The predicted statistical near-isotropy (redshift-space distortions are small outside the wedge) of the 21\,cm reionization signal makes the spherically-averaged power spectrum $P(k)$ an excellent statistic for making a high-sensitivity detection and connecting observations to theory and simulations. However, observationally there is a fundamental difference between the measurement of modes along the line of sight---which come from different observing frequencies---and modes perpendicular to the line of sight---which come from different baselines. Moreover, while the signal is expected to be nearly statistically isotropic, the foregrounds are not. Their smooth spectra make them distinguishable in cylindrically-binned Fourier space where the three Fourier wavenumbers $k_x$, $k_y$, and $k_z$ have been collapsed into two: $k_\| \equiv \left| k_z \right|$ and $k_\perp \equiv \sqrt{k_x^2 + k_y^2}$.

There are a number of approaches for forming power spectra from interferometric observations (e.g.\ \citealt{LT11, DillonFast, mapmaking, ShawPolarizedMMode, SimsBayesian, CHIPS, JacobsPipelines}). Following \cite{ModelingNoise}, we adopt the simplest approach which is to employ the delay-spectrum approximation \citep{AaronSensitivity}. In it, the cosmological 21\,cm power spectrum can be approximated to scale linearly with the mean amplitude square of the visibility Fourier transformed along the line of sight:
\begin{equation}\label{eq:power-approx}
   P(\mathbf{k} ) \approx \left(\frac{c^2}{2k_B^2\nu_0^2} \right)^2 \frac{X^2(\nu_0)Y(\nu_0)}{\beta_{pp}\Omega_{pp}} \left\langle \left|\widetilde{V} (\mathbf{u}, \eta)\right|^2\right\rangle,
\end{equation}
where $k_B$ is the Boltzmann constant, $\nu_0$ is the central frequency of the observation, and $\beta_{pp} \equiv \int df|b(\nu)|^2$ and $\Omega_{pp} \equiv \int d\Omega |B(\mathbf{\hat{r}})|^2$ are integrals over the squares of the bandpass and beam, respectively.  $X$ and $Y$ convert interferometric coordinates to comoving cosmological coordinates and are defined such that $2\pi(u,v,\eta)=(Xk_x, Xk_y, Yk_z)$. Here $u$ and $v$ are components of the baseline $\mathbf{b}$ in units of wavelengths and $\eta$ is the Fourier dual to frequency along the line of sight (i.e.\ at fixed |$\mathbf{u}$|).

The key idea behind the delay-spectrum approximation is to substitute the a delay-transformed visibility of single baseline, $\widetilde{V}_{ij}(\tau)$ for a the Fourier transform of gridded visibilities along the the line of sight, $\widetilde{V} (\mathbf{u}, \eta)$ \citep{AaronDelay}. $\widetilde{V}_{ij}(\tau)$, which is defined as
\begin{equation} \label{eq:delay-transform}
    \widetilde{V}_{ij}(\tau) = \int d\nu e^{2\pi i\nu \tau}V_{ij}(\nu),
\end{equation}
uses $\tau$ as the Fourier dual to frequency. The delay-spectrum approximation ignores the the frequency-dependence of the $\mathbf{u}$ modes probed by a single baseline. The approximation works best for short baselines and relatively small bandwidths. Most importantly, it allows us to estimate power spectra from single baselines. 

Another advantage of the delay-spectrum approach is that it gives us a straightforward way to understand the foreground wedge in terms of our visibilities. For a given baseline, the delay transform of the visibility spectrum maps sources based on their arrival times at the two antennas involved. Since the delay is at its maximum for a source on the horizon, there exists a ``horizon limit'' in delay space set by the baseline length \citep{AaronDelay}. As long as foregrounds are spectrally smooth, and thus produce narrow features in delay space, their power is restricted to lie inside the horizon. Instrumental spectral structure (due to miscalibration, e.g.) will scatter foreground power to high delays and thus contaminate high $k_\|$ modes of the delay spectrum.

In our simulations, we choose a band of 140 to 160\,MHz with 100 channels, which corresponds to a redshift range of $7.9<z<9.2$. 
This is similar to redshift range from which current EoR experiments aim to measure a single power spectrum (e.g.\ \citealt{JacobsPAPERMultiredshift}).
Because of the sidelobes produced by that finite bandwidth, we perform the discrete Fourier transform of our redundantly-calibrated visibilities with a Blackman-Harris windowing function \citep{BlackmanHarris}.
Following \cite{ModelingNoise}, we then form cylindrical power spectra by averaging together power spectra from different visibilities at a given $\left|\mathbf{u}\right|$. We then build up measurements at different $k_\perp$ by examining the delay-spectra from baselines of different lengths.

\begin{figure}
\includegraphics[width=\linewidth]{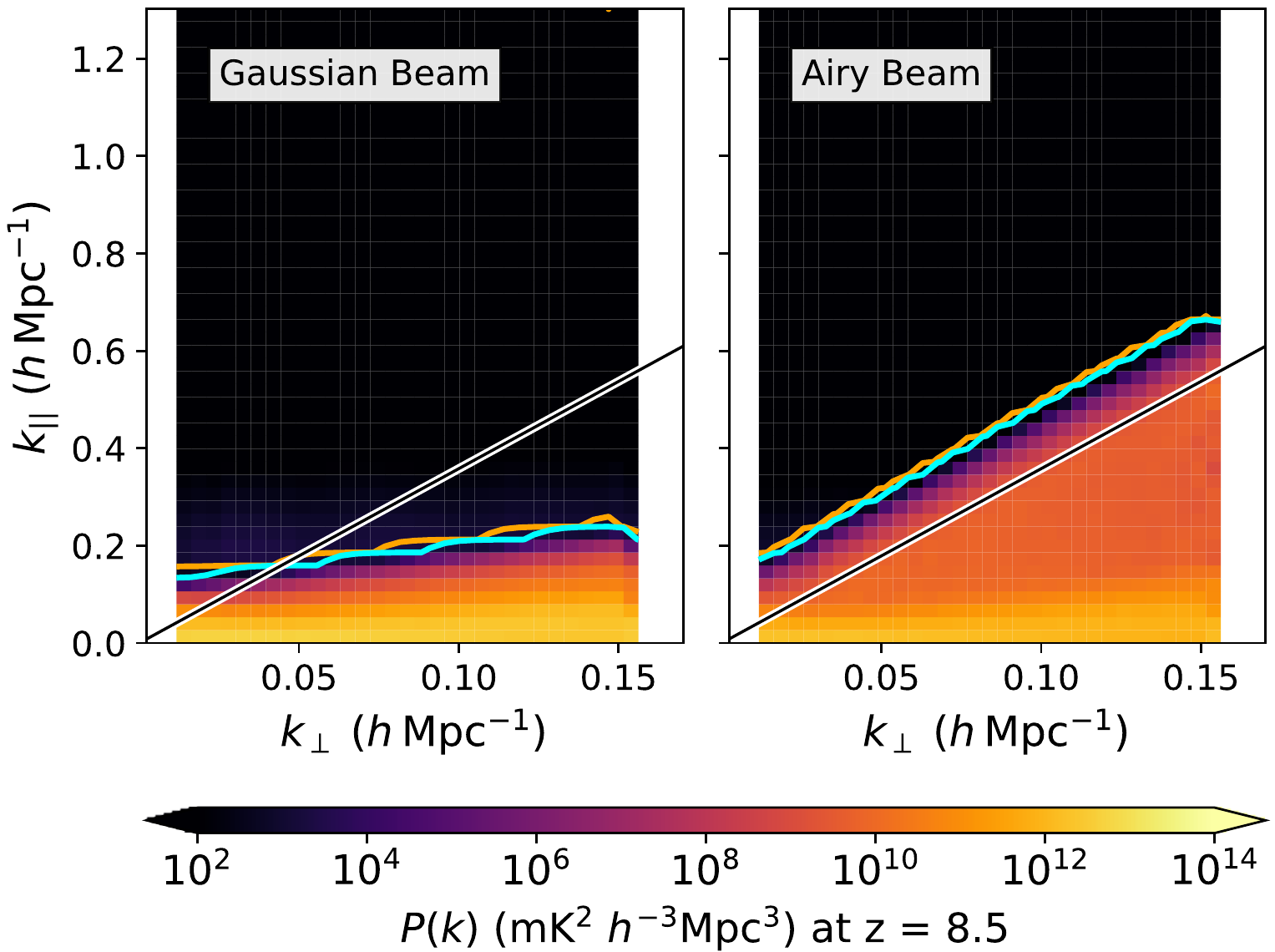}
\caption{Power spectra from a perfectly redundant array are shown with both Gaussian and Airy primary beams with the same FWHM. We also show the edge of the wedge corresponding to the delay at the horizon (black diagonal line) and contours where the foregrounds are equal to (cyan) and 10\% of (orange) a fiducial 21\,cm power spectrum (see Figure~\ref{fig:21cm_PS}). The Gaussian beam exponentially suppresses power from sources far from zenith, unrealistically eliminating the wedge. We adopt the more realistic Airy beam beam for all other simulations and power spectra in this work because it more faithfully approximates HERA's beam (or any beam from a dish) and lets us examine the effects of non-redundancy on delay power spectra in detail. The Airy beam shows contamination outside the wedge due to the finite bandwidth of the observation and the intrinsic width of the Blackman-Harris window in Fourier space.}
\label{fig:no_errors}
\end{figure}

We start by showing two cylindrically-binned power spectra with perfect calibration in Figure~\ref{fig:no_errors}. 
The straight diagonal line represents the horizon limit of the wedge in delay space. Due to their smooth spectral structure, foregrounds are mainly confined to the wedge, while the complex spectral structure of the 21\,cm brightness temperature fluctuations exhibits power at a range of delays that extends well outside the wedge. Because of the high dynamic range of the effect we are studying, we apply a Blackman-Harris window function to minimize foreground leakage due to the finite bandwidth of the simulation, reducing our effective bandwidth to approximately 10\,MHz. This keeps the EoR window mostly dominated by the 21\,cm signal, but produces some power just beyond the horizon limit, especially with an Airy beam model, due to the intrinsic width of the Blackman-Harris convolution kernel in Fourier space. Though foreground filtering may reduce this effect, we leave delay filtering (e.g.\ \citealt{AaronDelay}) to future work so as to isolate the impact of chromatic calibration errors.

Since the unphysical Gaussian beam exponentially suppresses sources close to the horizon, most of the wedge appears empty, thus hiding the effect of the most chromatic foregrounds. This creates the misleading impression that more modes are available to EoR science than is actually the case. The much more realistic Airy beam has much less suppression of foregrounds at the horizon. Realistic primary beams simply do not produce the $\sim$10$^5$ level of suppression needed to ignore sources near the horizon \citep{PoberWidefield}. We caution that adopting a Gaussian primary beam, especially one with a narrow FWHM, produces extremely unrealistic results in nearly any study of 21\,cm power spectrum estimation in the presence of foregrounds. We thus adopt the Airy beam (Equation~\ref{eq:airy_general}) for the rest of this work to realistically capture the effects of antenna-to-antenna variation in the sidelobes of the power pattern, which are both the most challenging to model \citep{NebenBeamforming} and the most important for understanding leakage outside the wedge.

\begin{figure}
    \centering
    \includegraphics[width=.85\linewidth]{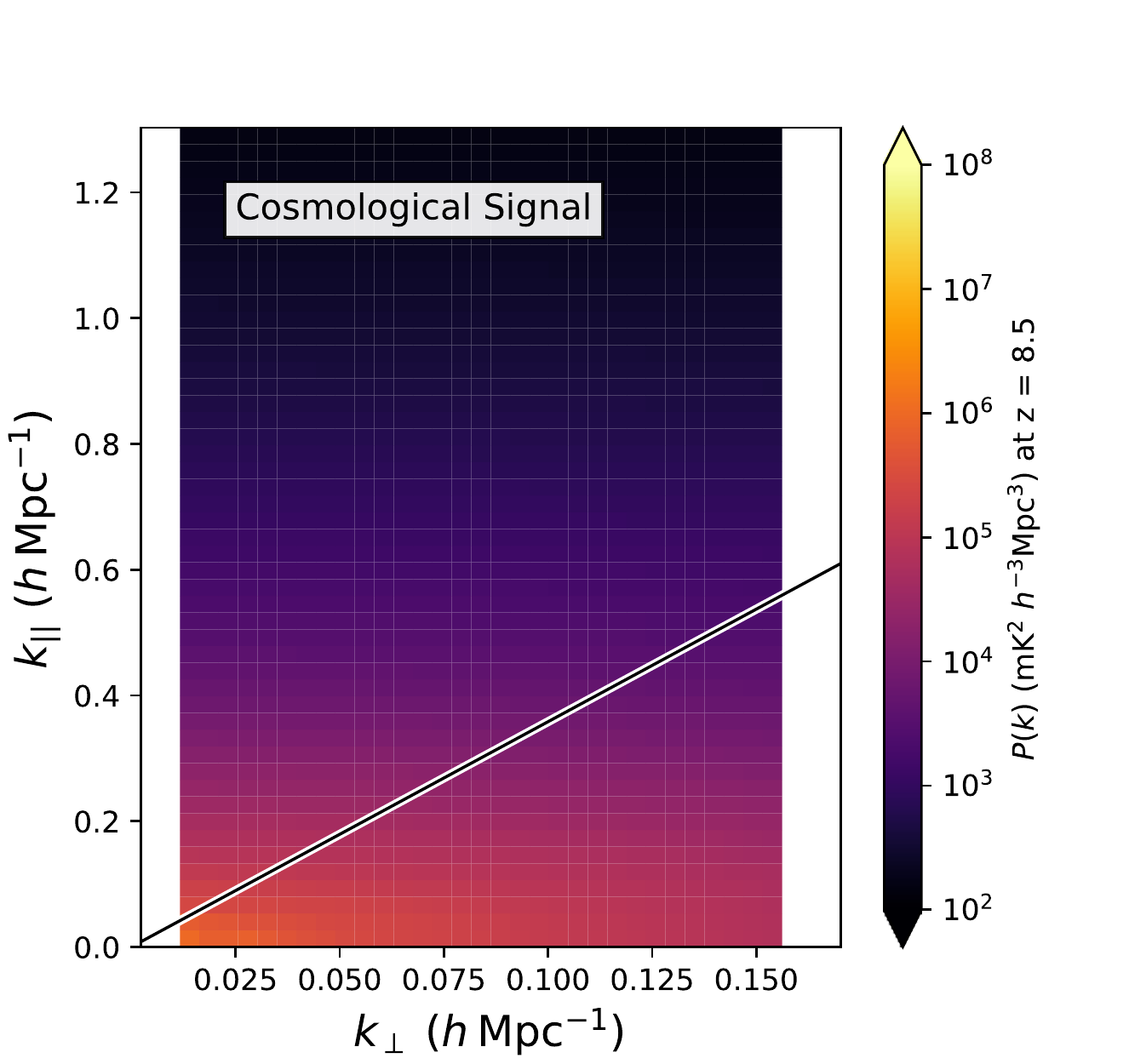}
    \caption{Cosmological signal from a {\tt 21cmFAST} \citep{Mesinger:2011} at $z\approx 8.5$, which is both the center redshift of our simulated observations and the redshift at which the universe is $\sim$50\% ionized with the fiducial set of astrophysical parameters. We assume that the cosmological signal is nearly statistically isotropic and thus only a function of $|\mathbf{k}|$, which is dominated by $k_\|$. A dense array concentrates sensitivity at low $k_\perp$, where one generally finds the largest 21\,cm signal magnitudes outside the wedge. Note that the colorscale in this figure and all subsequent analogous figures has been compressed relative to Figure~\ref{fig:no_errors} to show detail.}
    \label{fig:21cm_PS}
\end{figure}

To show which modes of the 21\,cm power spectrum are dominated by foreground bias, we overplot contours on all of our power spectra, which show where the simulated foregrounds have equal power to the 21\,cm signal (cyan) and where they have 10\% of the power of the 21\,cm signal (orange). As a representative cosmological signal (see Figure~\ref{fig:21cm_PS}), we use the popular three-parameter reionization model of \cite{Mesinger:2012} simulated by {\tt 21cmFAST}\footnote{\url{https://github.com/andreimesinger/21cmFAST}} \citep{Mesinger:2011} which includes the ionization efficiency $\zeta$, the mean-free path of UV photons in H{\sc ii} regions $\text{R}_\text{mfp}$, and the minimal virial temperature for dark-matter halos that host stars, $\text{T}_\text{vir}^\text{min}$. 
We set these parameters to fiducial values of $\zeta=20$,  $\text{R}_\text{mfp}=15\,\text{Mpc}$, and $\text{T}_\text{vir}=2 \times10^4$\,K which results in a 50\% reionized universe at $z \approx 8.5$ and a Thomson scattering optical depth to the CMB of $\tau \approx 0.08$. To reiterate, our simulations are foregrounds-only; we use a cosmological 21\,cm power spectrum for a quantitative comparison to foreground leakage but we do not simulate visibilities from a realization of this power spectrum. Nor do we take into account the mixing of cosmological modes introduced by the delay-spectrum approximation because that effect is marginal compared to the foreground leakage.

\begin{figure*}
\includegraphics[width=\textwidth]{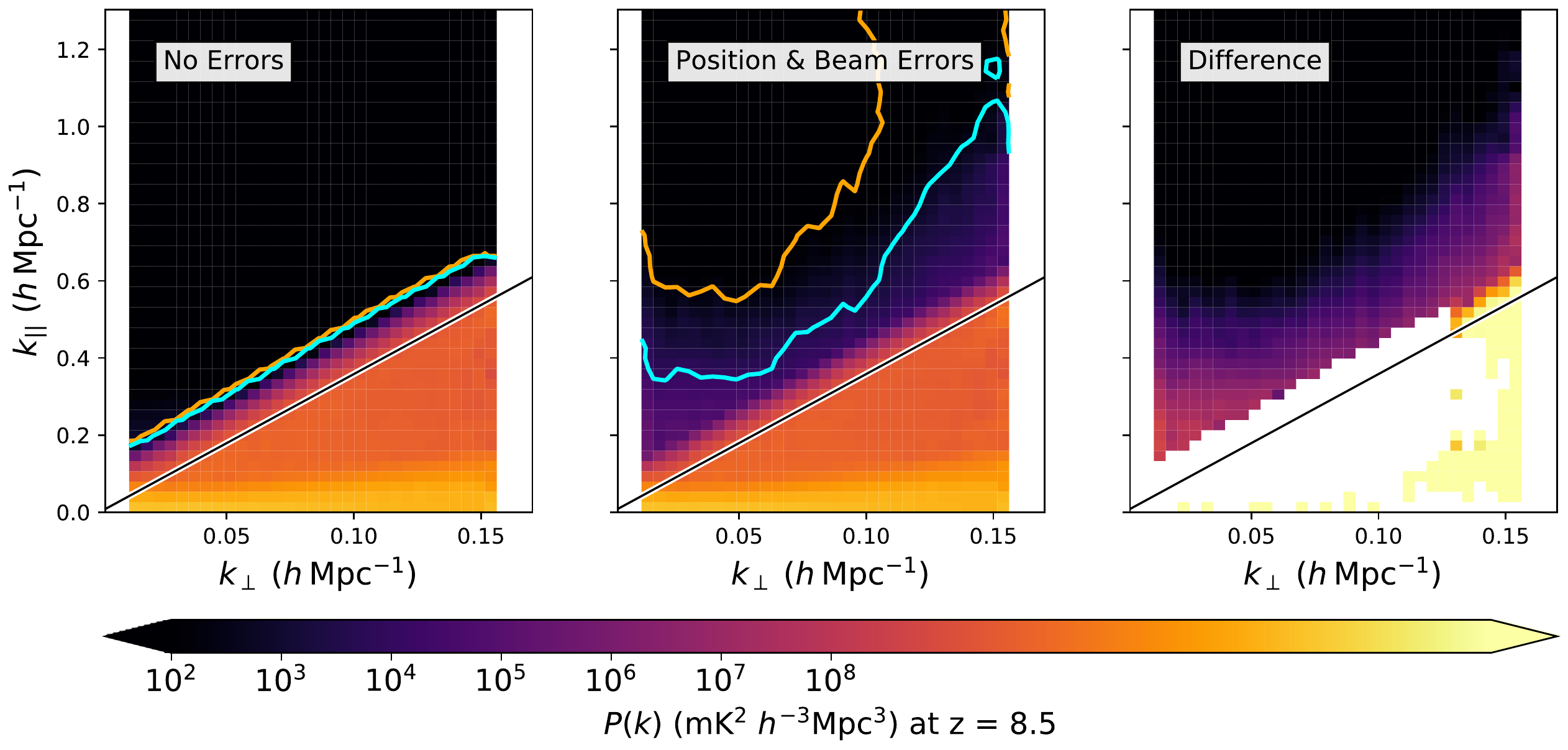}
\caption{Foreground power spectra with and without the effect of simulated antenna-to-antenna variation on redundant-baseline calibration. Non-redundancy produces considerable leakage of foreground power from the wedge (below the black diagonal line) into the window. Though the leaked power is small compared to the total foreground power, foregrounds are $\sim$10$^5$ brighter than the fiducial cosmological signal (see Figure~\ref{fig:21cm_PS}). To illustrate the importance of this effect, we overplot contours where the foreground bias is equal to (cyan) and 10\% of (orange) the cosmological signal. Compared to the case with no redundancy errors, our fiducial level of position and beam errors leads to contamination dominating a considerable fraction of the EoR window.  On the right, we show the difference in power between the left-hand and middle panels, which clearly shows a transfer of power from the wedge (which is negative and thus white) to the window.}
\label{fig:mid_errors}
\end{figure*}

\begin{figure*}
\includegraphics[width=\textwidth]{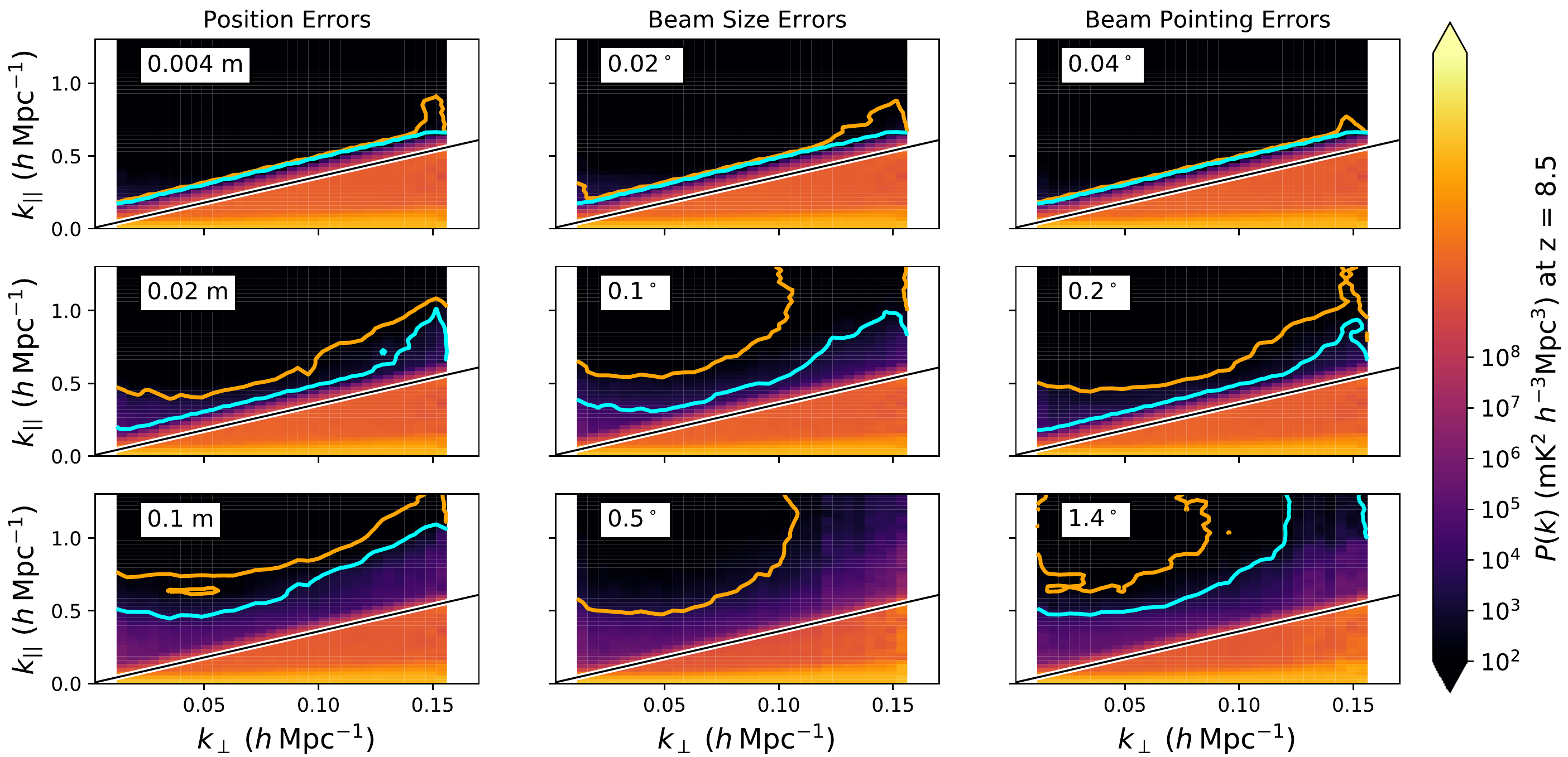}
\caption{Power spectra broken down by error type (columns) and error level (rows) show that increasing error level creates larger and larger contaminated regions of the EoR window. We have simulated power spectra for each error type (see Table~\ref{errors_table}), individually assessing antenna position errors (left column), beam size/shape errors (middle column), and beam pointing errors (right column). The effect for our low error level (top row), with it's $\sim$1\% visibility standard deviation within redundant groups, is minimal. However, the fiducial and high error levels produce systematic errors in much of the EoR window that dominates over our fiducial signal. Note that our choice of color scale is compressed therelative to the full dynamic range for the foregrounds (roughly 10 order of magnitude in power). This better shows detail outside the wedge, the primary region of interest, but saturates the color scale inside the wedge (see, Figure~\ref{fig:no_errors} for an unsaturated wedge).}
\label{fig:errors_chart}
\end{figure*}

Redundancy errors impede separation of cosmological signal from foregrounds in Fourier space by imparting spectral structure to the intrinsically smooth foregrounds and moving them into higher Fourier modes that were previously signal-dominated. Figure~\ref{fig:mid_errors} shows that fiducial errors (see Table~\ref{errors_table}) lead to significant contamination of the EoR window. 
Power leaves the wedge (shown in white in the right-hand panel of Figure~\ref{fig:mid_errors} where the change in power is negative) and moves to higher delays, obscuring the 21\,cm signal. This illustrates how long and short baselines are interlinked during calibration, allowing spectrally-unsmooth errors from long baselines to affect short ones. Most importantly, the region with the highest signal-to-noise, the bottom-left corner, is now completely dominated by foreground bias. To break down this effect further, in Figure~\ref{fig:errors_chart} we compare power spectra of simulations with a single type of redundancy error (position, beam size, and beam pointing) at low, fiducial, and high error levels.
With low levels of error (top row), the effect on the power spectra is minimal. However, at the fiducial (middle row) and high levels of error (bottom row), power spreads out of the wedge and covers more of the EoR window. For comparable levels of visibility variance introduced by these antenna-to-antenna variations (Figure~\ref{fig:STDEV}), the calibration with non-redundancy produces similar levels of EoR window contamination for all three error types, though perhaps beam size/shape errors are slightly more deleterious than the other two types. Regardless, it is clear that doing 21\,cm cosmology in the window will require an improvement in our approach to calibration. 

\begin{figure*}
\includegraphics[width=\textwidth]{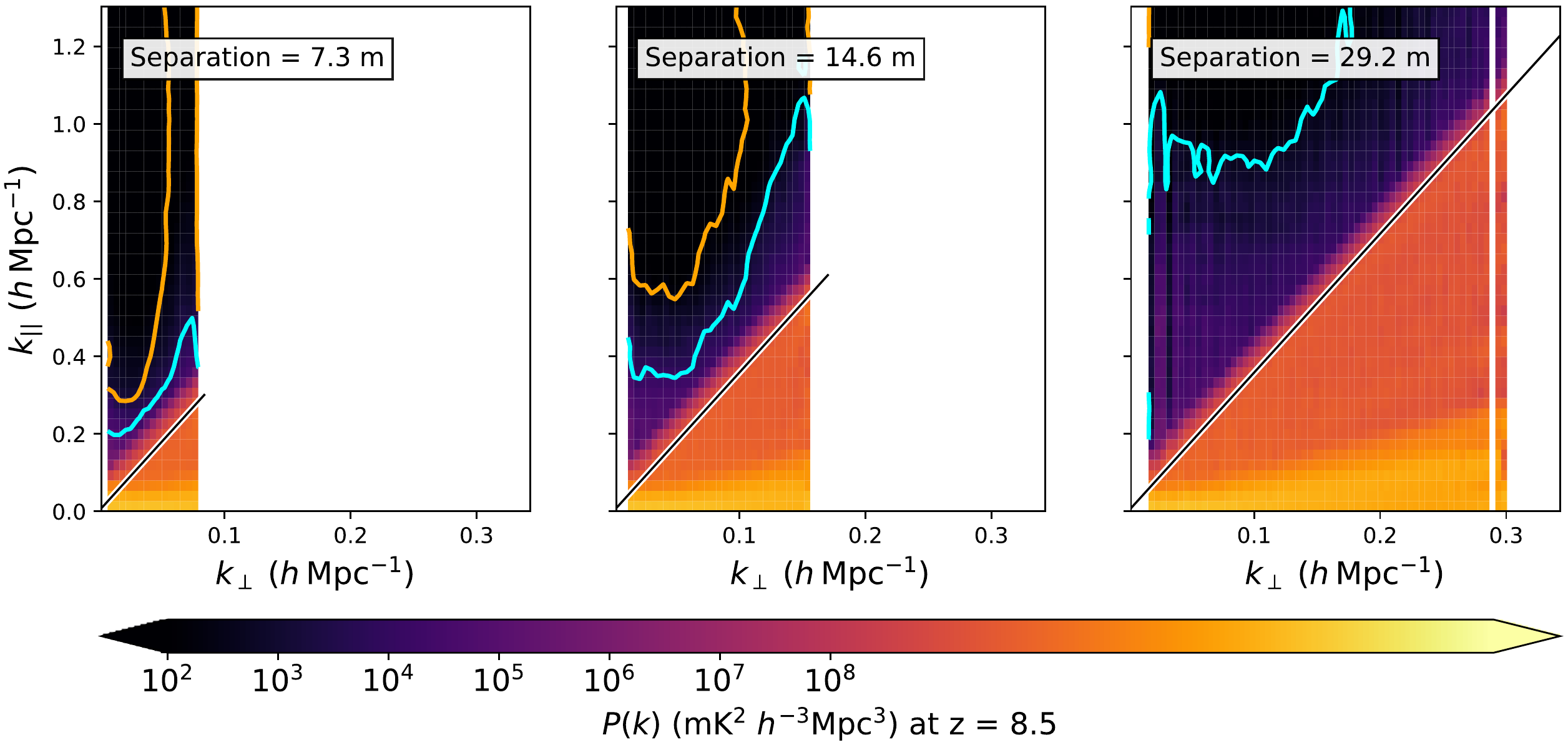}
\caption{Introducing non-redundancies at the fiducial error level, we produce power spectra from three arrays which only differ by their antenna separation (and thus overall size) but still have analogous sets of redundant baselines. The standard HERA core is shown in the middle panel. Larger arrays have longer baselines, which cover higher values of $k_\perp$ (note the wider range of $k_\perp$ than in the above figures). During calibration, these long baselines leak the spectral structure of their visibilities to short baselines through gains. This causes the wedge to increase in area and the contamination to leak higher into the EoR window across all values of $k_\perp$. The overall size of the contaminated area above the wedge, as well as its extent in $k_\|$, scale with the length of the longest baseline.}
\label{fig:size}
\end{figure*}

We have argued that the key to understanding the effect of non-redundancy is the leakage of spectral structure from long baselines to short baselines via chromatic gain errors. To see the impact of long baselines more directly, we repeat the experiment from Section~\ref{sec:non-redcal} of looking at how the power spectrum is affected when we look at alternate 320-element HERA cores with different inter-element spacing and thus different longest baselines. Using the same (unphysical) half- and double-sized arrays we described in Figure~\ref{fig:gains}, we plot these power spectra in Figure~\ref{fig:size}. The half-sized array has less $k_\perp$ coverage because its baselines are shorter while the double-sized array covers higher values of $k_\perp$ than the standard HERA core. 
Because the wedge extends to higher values of $k_\|$ for the longest baselines of the larger arrays, the leakage due to redundancy errors on the short baselines also extends to higher values of $k_\|$. Ultimately, including long baselines in the calibration---and putting them on equal footing with short baselines---means redundancy errors on long baselines are allowed to affect the calibrated visibilities of short baselines (and the reverse as well, though that is far less impactful). 

We now turn to a modified calibration scheme designed to isolate short baselines from long ones and limit the spectral structure of the inevitable calibration errors that arise due to antenna-to-antenna variation whose precise magnitude is not well known.


\section{Mitigating Redundancy Errors} \label{sec:mitigating}

As the bias to the power spectrum due to redundancy errors does not integrate down with repeated observations of the same field, it is necessary to find a way to calibrate with minimal extra power above the horizon limit if we want to maximize the area of Fourier space available to cosmology. In Section~\ref{sec:cutoff} we adapt the weighting techniques of \cite{ModelingNoise} to propose the use of baseline length cutoff for redundant calibration and discuss the details of implementing this weighting scheme. Then in Section ~\ref{sec:cutoff_results}, we compare results and power spectra from different implementations and show that the chromatic effect on the EoR window of our fiducial level of non-redundancy demonstrated in Section~\ref{sec:simulations} can be largely eliminated with an appropriate baseline-length-dependent weighting scheme.

\subsection{Redundant calibration with baseline-length weighting}\label{sec:cutoff}

We saw in Figure~\ref{fig:gains} that the the spectral structure in gain errors increased with longer maximum baseline length and that this effect leads to a more contaminated EoR window (Figure~\ref{fig:size}). It follows then that we need a calibration approach that isolates long baselines by suppressing their contribution to the gains, thus keeping them from spreading chromatic errors to short baselines. As an alternative to using the inverse noise covariance to weight visibilities in Equation~\ref{eq:x_with_weights}, we instead propose assigning binary weights based on baseline length, imposing a ``cutoff` on which baselines are included in gain calibration, i.e.\ picking $\mathbf{W}$ to be a diagonal matrix with zeros for long baselines and ones for short baselines. This technique is a modification of the weighting scheme proposed in \cite{ModelingNoise}, which instead proposed a Gaussian weighting of visibilities as a function of their length. 

Using binary weights is equivalent to calibrating with only a subset of the data. First we use redundant-baseline calibration to simultaneously find all the antenna gains as well as the unique visibility solutions for short baselines. Then we can use the gain solutions to calibrate the visibilities of the long baselines and average over redundant baselines to get an estimate of the remaining unique visibility solutions. In this way, the entire array is calibrated but the gain solutions are isolated from the long baselines. 

One advantage of giving zero weight to long baselines (rather than a small weight as a Gaussian scheme does) is that that this reduces the computational complexity of calibration. The most expensive step in redundant calibration is the inversion of $\mathbf{A}^\intercal \mathbf{W}\mathbf{A}$ in each iteration of \texttt{lincal}. This scales as $\mathcal{O}(N^3)$, where $N$ is the number of variables being solved for simultaneously, i.e.\ the number of antennas plus the number of unique baselines included in the calibration. HERA's core of 320 antennas has 1501 unique baselines, so without any cutoff, $N$ is dominated by the number of unique baselines. 

\begin{figure}
\includegraphics[width=\linewidth]{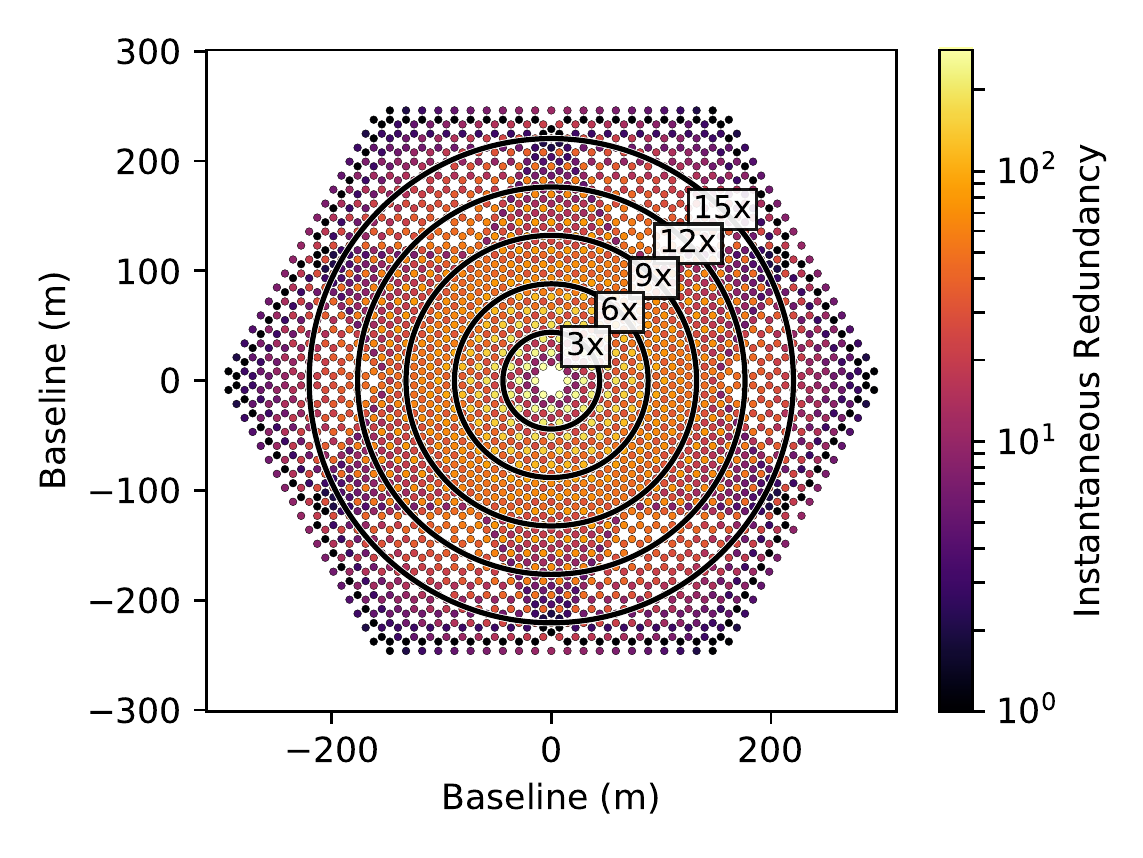}
\caption{The HERA core's unique baselines, colored by their instantaneous redundancy (i.e.\ the number of total baselines measuring that unique baseline). The baseline length cutoffs we use in this work (3, 6, 9, 12, and 15) are marked with a circle encompassing all unique baselines still included.}
\label{fig:redundancy}
\end{figure}

In Table \ref{cutoffs_table}, we list the statistics for a sample of different baseline-length cutoffs, including the smallest cutoff that is still redundantly calibratable (the 27 shortest unique baselines).\footnote{Smaller cutoffs introduce additional degeneracies beyond the four inherent to redundant-baseline calibration \citep{DillonPolarizedRedundant}.}
\definecolor{Gray}{gray}{0.9}
\setlength{\tabcolsep}{10pt} 
\renewcommand{\arraystretch}{1.2} 
\begin{table*}
\centering
    \begin{tabular}{cccc}
         \hline
            \shortstack{\textbf{Baseline Cutoff In Units}\\\textbf{Of The Shortest Baseline}} & \shortstack{\textbf{Total Baselines Included}\\\textbf{(Percentage)}} & \shortstack{\textbf{Unique Baselines Included}\\\textbf{(Percentage)}} & 
            \shortstack{\textbf{Median Thermal Gain Error}\\\textbf{In Units Of Visibility N/S}} \\
        \hline
        \rowcolor{Gray}
            Minimum  & 2545 (5.0\%) & 27 (1.8\%) & 0.246 \\
        \hline
            3x & 4595 (9.0\%) & 45 (3.0\%) & 0.183 \\
        \hline
        \rowcolor{Gray}
            6x & 14362 (28\%) & 189 (13\%) & 0.107 \\
       \hline
            9x & 27105 (53\%) & 435 (29\%) & 0.079 \\
        \hline
        \rowcolor{Gray}
            12x & 39075 (77\%) & 751 (50\%) & 0.065 \\            
       \hline
            15x & 47167 (92\%) & 1086 (72\%) & 0.058 \\
        \hline
        \rowcolor{Gray}
            No Cutoff & 51040 (100\%) & 1501 (100\%) & 0.057 \\
        \hline
    \end{tabular}
    \caption{ The number of baselines included, the number of unique baselines, and their respective percentages of the total for different cutoff levels (expressed as multiples of the minimum antenna separation) is listed above. The minimum number of unique baselines that does not introduce additional degeneracies is 27. HERA can be redundantly calibrated using just $5\%$ of all its baselines. We also provide an estimate for the median gain errors due to thermal noise, normalized by the noise-to-signal ratio on visibilities.}
    \label{cutoffs_table}
\end{table*}
For simplicity, we use baseline-length cutoffs that are integer multiples of the minimum antenna separation, but the procedure can be generalized to any cutoff criterion above the minimum. To help visualize these cutoffs, we marked them on a scatter plot of all of the HERA core's unique baseline vectors in Figure~\ref{fig:redundancy}.

Due to HERA's dense configuration, the shortest baselines are also the most redundantly sampled. This means that a relatively small fraction of the unique baselines can be included without throwing out much information. Generally speaking, thermal noise on gain solutions contributes very little to the noise on calibrated visibilities. However, when using a relatively small subset of the available information for calibration, one might worry that we risk gain errors due to thermal noise alone (rater than non-redundancy) that contribute significantly to the noise on calibrated visibilities. 

Following \citet{RedArrayConfig}, we estimate the thermal noise error on our gains in the simplified scenario where all visibilities measure the sky with same signal-to-noise ratio (S/N). In this case, the expected thermal gain covariance is given by.
\begin{equation}
\mathbf{\Sigma} \approx \frac{1}{\left(\text{S/N}\right)^2} \left[\mathbf{A}^\dagger\mathbf{A}\right]^{-1}
\end{equation}
Reading gain variances off the diagonal of this matrix, which we compute using \texttt{logcal} \citep{redundant,RedArrayConfig} and report in Table~\ref{cutoffs_table} in units of N/S. To first order, these thermal gain errors produce extra thermal noise in calibrated visibilities equal to noise on the gains of the two antennas involved. Thus, even even for very small cutoffs, this effect is subdominant to the measured noise on the visibilities themselves. Similarly, in the context of sky-based calibration, \citet{ModelingNoise} also find little impact of down-weighting long baselines on the thermal noise of calibrated visibilities. Therefore one should generally pick a cutoff based its impact on chromatic systematic errors.

\begin{figure}
\includegraphics[width=\linewidth]{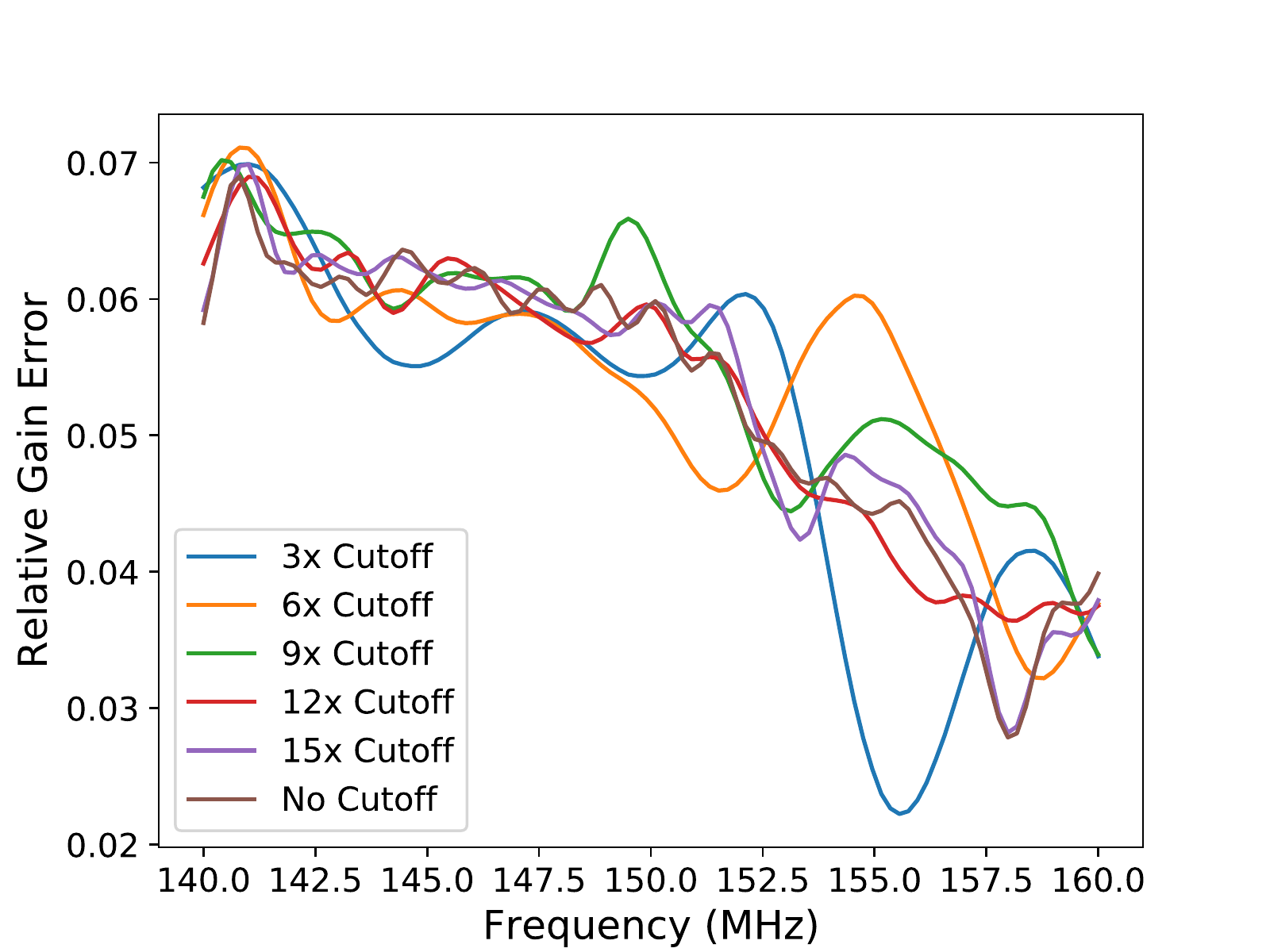}
\caption{Here we show the relative error on gain solutions for an antenna in the Northwest corner of the array with our fiducial level of non-redundancy. While the simulated visibilities are identical and noise-free in all cases, restricting the baseline-length cutoff used in calibration decreases the spectral structure in the derived gains. We examine baseline-length cutoffs at 3, 6, 9, 12, and 15 times 14.6\,m antenna separation. Lower cutoffs prevent the longer baselines' spectral structure from affecting the gain solutions, as we saw in Figure~\ref{fig:gains}. This trend holds for antennas throughout the array.}
\label{fig:gain_error_subset}
\end{figure}

To begin investigating the impact of a baseline-length cutoff, we return to the question of its qualitative impact on the spectral structure of gain errors. Analogous to Figure~\ref{fig:gains}, we show in Figure~\ref{fig:gain_error_subset}, the difference between simulated true gains and calibrated gain solutions with different baseline-length cutoffs for a representative antenna in the HERA core. 
A high cutoff means only the longest baselines are excluded, while a low cutoff excludes all but the shortest baselines. Isolating longer baselines this way prevents their spectral structure from spreading to shorter baselines through gains. Figure~\ref{fig:gain_error_subset} lends credence to our hypothesis; as we restrict our calibration to shorter and shorter baselines, we see less and less spectral structure. This is imperative if we want to keep foregrounds isolated in the wedge, which we will now investigate.

\subsection{Impact of a baseline-length cutoff on the power spectrum} \label{sec:cutoff_results}

\begin{figure*}
\includegraphics[width=\textwidth]{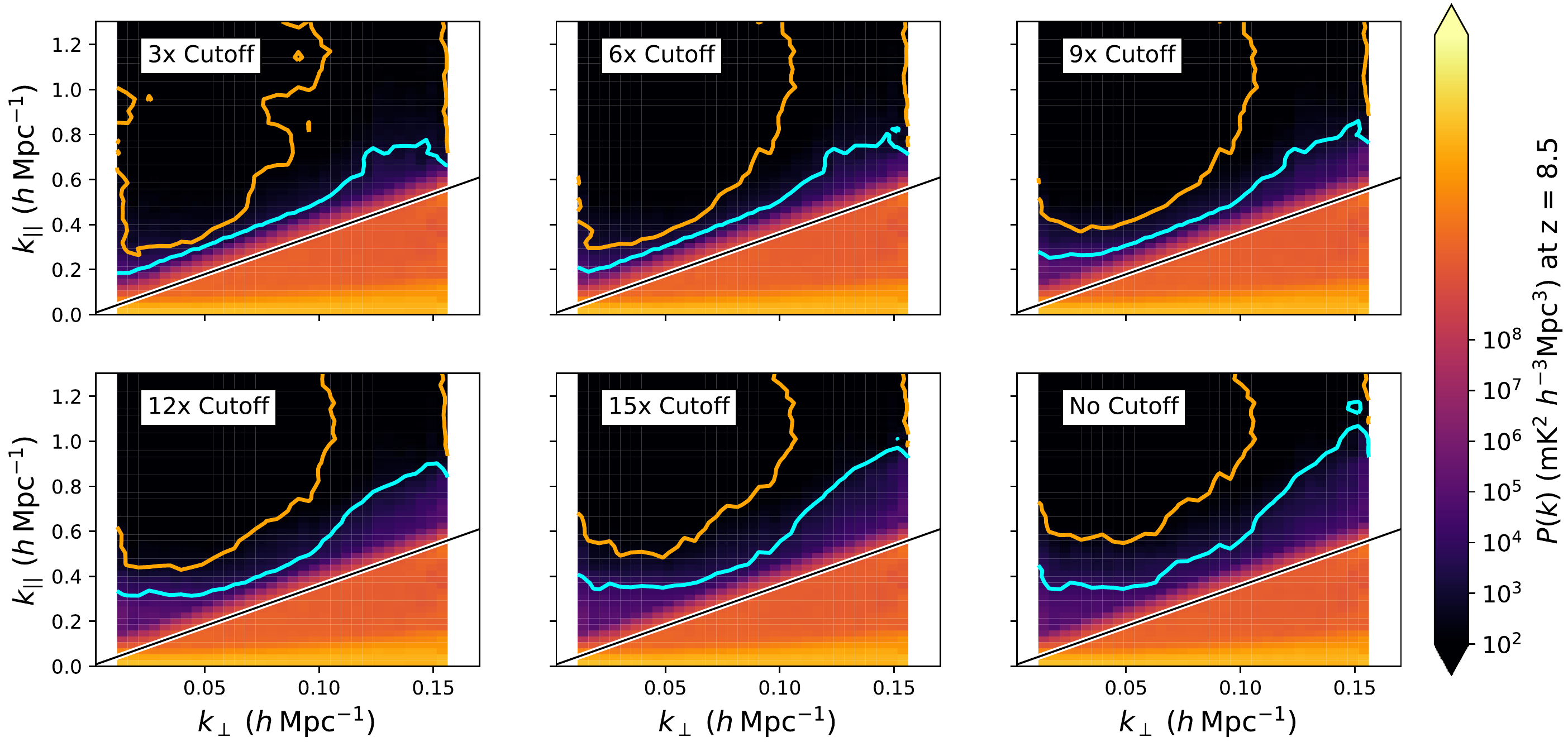}
\caption{The effect of our fiducial redundancy errors on the EoR window is increasingly mitigated as we impose stricter baseline-length cutoffs on our redundant calibration. We show cutoffs at 3, 6, 9, 12, and 15 times the shortest baseline, as well as calibration without a cutoff (which is identical to the middle panel of Figure~\ref{fig:mid_errors}). Excluding long baselines from the calibration prevents their intrinsic spectral structure from contaminating the short baselines---those most sensitive to the 21\,cm signal---via the gains. At some point, likely between $3\times$ and $6\times$, that effect is outweighed by the increased variance due to a relatively low number non-redundant visibilities combined to estimate the gains.}
\label{fig:cutoffs}
\end{figure*}

Since measurement of the fluctuations in the cosmological signal will occur through power spectra, the best test of the efficacy of calibration with a baseline-length cutoff is to analyze power spectra.
In Figure~\ref{fig:cutoffs}, we compare five different baseline-length cutoffs to a calibration with no cutoff in order to see how much of the foreground leakage created by our fiducial redundancy errors they remove.
We use the evenly-spaced cutoffs from Table~\ref{cutoffs_table} ranging from 3 to 15 times the shortest baseline in the array. In general, we find that smaller cutoffs perform better. 

The most effective cutoffs, $3\times$ and $6\times$, perform similarly, largely pushing down foreground leakage back to where it was observed in our simulation with no antenna-to-antenna variation (see Figure~\ref{fig:no_errors}), which we attribute to finite bandwidth effects and the Blackman-Harris window. While low levels of foreground contamination persist in large portions of the EoR window, the highest signal-to-noise region (bottom-left) is the least contaminated. By comparison, power spectra with large or no cutoffs progressively show larger and larger regions of the window that are biased at levels exceeding the 21\,cm signal. This clearly demonstrations that the inclusion of even moderately-long baselines in calibration can drastically affect the foreground contamination on all baselines, spreading power above the wedge and into the window. 

Figure~\ref{fig:cutoffs} raises an important question: how do we pick the optimal cutoff? Obviously the best cutoff is fairly small, but there exist trade-offs that should be considered. Larger cutoffs include more information, which means that errors due to noise (see Table~\ref{cutoffs_table}) and due to antenna-to-antenna variation average down a bit better. Smaller cutoffs have the least contamination of the EoR window, especially in the all-important bottom-left corner (low $k_||$ and $k_\perp$). Likely, the optimal cutoff also depends on the level of non-redundancy, the foreground and instrument model, and the techniques of foreground mitigation employed (e.g.\ foreground subtraction or delay filtering). A systematic study of this effect is beyond the scope of this paper; our aim is merely a proof of concept.

\begin{figure*}
\includegraphics[width=\textwidth]{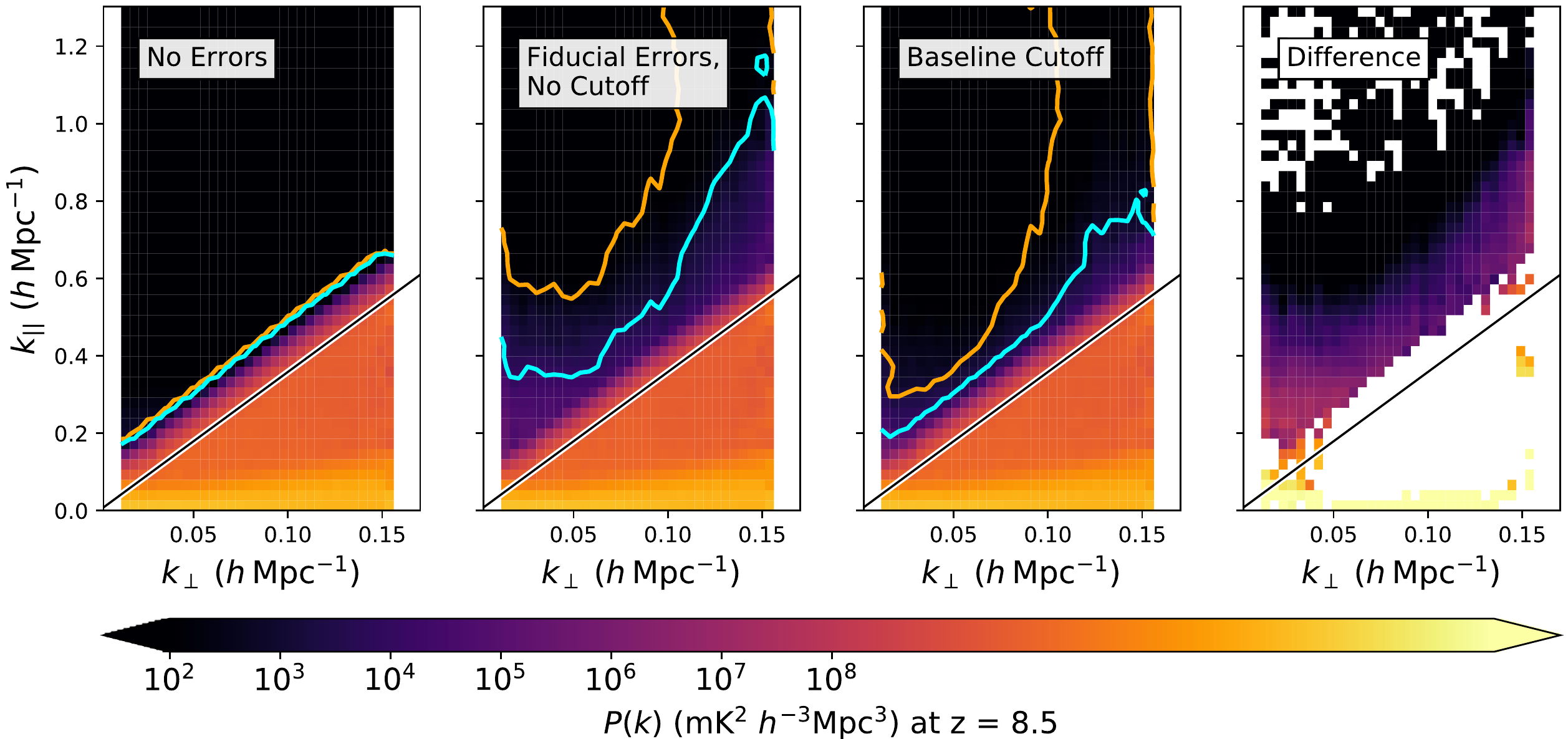}
\caption{Reweighting our data by imposing a $6\times$ baseline-length cutoff (third panel) recovers much of the EoR window that was previously contaminated by spectral structure in the calibration errors (second panel) without the cutoff. In fact, most of the EoR window accessible to 21\,cm cosmology in the perfectly calibrated case (first panel) now signal-dominated again. The difference between the second and third panels, shown in the last panel, shows how leaked foregrounds are transferred out of the window and into the wedge, which is negative (and thus white) in the difference. In this simulation, we used the fiducial non-redundancy level (Table~\ref{errors_table}) and a baseline-length cutoff of 6 times the 14.6\,m antenna separation.}
\label{fig:cutoff_v_uw}
\end{figure*}

For a more detailed demonstration of the effectiveness of our baseline cutoff, we pick $6\times$ (top middle of Figure~\ref{fig:cutoffs}) as \textit{close enough} to optimal; it is clearly better than $9\times$ at removing low $k_\perp$ foreground bias but shows less structure in the signal-to-bias contours than $3\times$ (likely due to more visibilities with uncorrelated non-redundancy averaging down incoherently). In Figure~\ref{fig:cutoff_v_uw}, we compare our $6\times$ result a simulation with no errors and to the result with fiducial levels of non-redundancy from Section~\ref{sec:simulations} but with no baseline-length cutoff.
Clearly, our reweighted power spectra (third panel) recover much of the EoR window that was lost to foreground bias without the cutoff (second panel). Encouragingly, the reweighted power spectrum appears quite similar to the perfectly-redundant power spectrum shown in the left panel, except for a small amount of increased contamination for the highest bins in $k_\perp$. 

Finally, it is important to note that though our weighting scheme reduces power in the EoR window, it does not in general produce more accurate calibration solutions. In fact, if we look at the right-hand panel of Figure~\ref{fig:cutoff_v_uw}, we see that while foreground bias went down in the window, it actually went up in the wedge (white region). In other words, we did not reduce our calibration errors; we merely contained them. To work within the EoR window, we do not necessarily need calibration solutions that are accurate to one part in $10^5$---these calibration solutions are inaccurate at the $1$-$10\%$ level---we just need to make sure that our errors are very spectrally smooth.


\section{Conclusion}
Foregrounds that are roughly 10$^5$ times brighter than the cosmological signal pose a fundamental challenge for 21\,cm intensity mapping. While spectral smoothness is key to foreground separation, an interferometer is an inherently chromatic instrument. It naturally takes foreground power and spreads it out into a region of 2D Fourier space called the ``wedge,'' outside of which we have a putatively clean ``window'' to measure the cosmological signal. And yet, the extreme dynamic range requirements of 21\,cm cosmology makes the wedge/window distinction vulnerable to calibration errors. A small amount of chromatic miscalibration, when multiplied by the overwhelmingly bright foregrounds, leaks power into previously clean regions of Fourier space and biases the measurement. In the context of sky-based calibration, \citet{BarryCal} and \citet{ModelingNoise} found that even very small errors in one's radio source catalog can lead to leaked foreground power in the window at a level well above the signal.

Following that work, we examined in this paper how real-world challenges can complicate our attempts to calibrate using the self-consistency of nominally redundant baselines. Using the Hydrogen Epoch of Reionization Array (HERA) as a worked example, we investigated the effects on the power spectrum of a calibration scheme that assumes perfect redundancy of visibility measurements while introducing antenna-to-antenna variation of positioning, beam size/shape, and beam pointing that broke that assumption. In Section~\ref{sec:simulations} we found an analogous effect that of \citet{BarryCal} and \citet{ModelingNoise}; non-redundancy introduced chromatic errors in our calibration solutions with spectral structure up the scale of the longest (and thus most chromatic) baselines included in the calibration. With reasonable levels of antenna-to-antenna variation, this produced foreground bias in the EoR window that significantly shrank the region accessible to a cosmological measurement, diminishing the ultimate sensitivity of HERA or any similar instrument.

However, inspired by the baseline-length weighting introduced by \citet{ModelingNoise}, we investigated the effects of imposing a baseline-length cutoff on which visibilities to include in our redundant-baseline calibration system of equations. We found in Section~\ref{sec:mitigating} that by limiting ourselves to the shortest baselines, we eliminated much of the spectral structure in our calibration errors and thus in the calibrated visibilities on our shortest baselines, which are both the most sensitive to the 21\,cm signal and the least contaminated by the wedge.

This work remains merely a proof of concept. Our sky model does not capture the full complexity of the polarized point sources and diffuse galactic emission in the real radio sky. Our parameterization of non-redundancy, while powerful, is ultimately a major simplification. Our Airy beam, while certainly much more faithful than a Gaussian beam, still deviates significantly from real HERA beams. We make no attempt to use external information about our instrument---if we knew how our antennas deviated from the ideal, perhaps we could use that to improve our weighting scheme or our calibration. Likewise, we do not explore the full parameter space of possible visibility weightings, adopting instead the simplest technique. Our simulated gains are assumed to be spectrally smooth, though realistic gains have spectral structure---at least on large scales. And of course, there are other ways to reduce the degrees of freedom in a calibration solution to ensure smoothness, including time averaging \citep{BarryCal} and low-pass filtering. These complications and extensions are left for future work.

That said, the fact that the effect of non-redundancy can be largely mitigated means that redundant-baseline calibration remains a powerful, albeit incomplete, technique for meeting the exacting demands of 21\,cm cosmology. While the non-redundancies intrinsic to any real instrument will produce calibration errors, not all errors are created equal. Our strategy of baseline-length weighting is fundamentally one of managed ignorance; redundant-baseline calibration will produce errors, but with a careful reweighting, we can avoid errors that prevent us from measuring the cosmological 21\,cm signal. 


\section*{Acknowledgements}
The authors wish to thank Zaki Ali, Judd Bowman, Philip Bull, Carina Cheng, Dave DeBoer, Deepthi Gorthi, Nicholas Kern, Wenyang Li, Adrian Liu, and Jonathan Pober for valuable discussions.
This work is supported by the National Science Foundation under grants \#1440343 and \#1636646, the Gordon and Betty Moore Foundation, and with institutional support from the HERA collaboration partners. JSD gratefully acknowledges the support of the NSF AAPF award \#1701536 and the Berkeley Center for Cosmological Physics. ARP acknowledges support of NSF CAREER award \#1352519 and the University of California Office of the President Multicampus Research Programs and Initiatives through award MR-15-328388 as part of the University of California Cosmic Dawn Initiative.

\bibliographystyle{mnras}
\bibliography{refs}

\label{lastpage}
\end{document}